\documentclass[a4paper,twoside]{article}

\usepackage{graphicx,fullpage}
\usepackage{bm}
\usepackage{amsfonts,amsmath,amscd,latexsym}
\usepackage{natbib}
\bibpunct{(}{)}{,}{a}{,}{,}

\newcommand\locsection{\subsection}

\newcommand{\autfont}{\fontfamily{cmr}\fontshape{it}
  \fontseries{m}\fontsize{11}{12}\selectfont}
\newcommand{\aut}[1]{\vspace*{2mm}\centerline{{\autfont\scshape #1 \hfil}}}

\newcommand{\ted}{\,\text{d}}
\newcommand{\BE}{\mathbb{E}}
\newcommand{\BP}{\mathbb{P}}
\newcommand{\ev}{\mathfrak{Z}}

\begin{document}

\title{\bf On Particle Learning\footnote{N.~Chopin and C.P.~Robert are partly supported by the 2007--2010 grant ANR-07-BLAN-0237-01 ``SP Bayes".
J.-M.~Marin and C.P.~Robert are partly supported by the 2009--2012 grant ANR-09-BLAN-0218 ``Big'MC".
Robin Ryder is funded by a postdoctoral fellowship from the Fondation des Sciences Math\'ematiques de Paris.
Christian Sch\"afer is supported by a PhD grant from CREST.}}

\author{\aut{Nicolas Chopin$^1$, Alessandra Iacobucci$^2$, Jean-Michel Marin$^{1,3}$,}\\
\aut{Kerrie L.~Mengersen$^4$, Christian P.~Robert$^{1,2,4}$,}\\
\aut{Robin Ryder$^{1,2}$, and Christian Sch\"afer$^{1,2}$}\\
{$^1$CREST, Paris}\hglue 1truecm 
{$^2$Universit\'e Paris-Dauphine, CEREMADE}\\
{$^3$IM3, Universit\'e Montpellier 2}\hglue 1truecm 
{$^4$Queensland University of Technology}
}

\maketitle

\begin{abstract}
This document is the aggregation of several discussions of \cite{lopes:carvalho:johannes:polson:2010} we
submitted to the proceedings of the Ninth Valencia Meeting, held in Benidorm, Spain, on June 3--8, 2010,
in conjunction with Hedibert Lopes' talk at this meeting.
The main point in those discussions is the potential for degeneracy in the particle learning methodology,
related with the exponential forgetting of the past simulations. We illustrate the resulting difficulties
in the case of mixtures.
\end{abstract}

\noindent{\bf Keywords:}
Attrition; degeneracy; evidence; importance sampling;  Marginal likelihood; Markov chain Monte Carlo; mixtures of
distributions; particle filter; sequential sampling; simulation.

\section{The case of mixtures (Mengersen, Iacobucci and Robert)}
\bigskip
In this discussion, we primarily consider the performances of the particle learning (PL) technique of 
\cite{lopes:carvalho:johannes:polson:2010} in the specific case of mixtures of distributions.

\locsection{Particle learning}
Reminiscing similar remarks made during Professor Polson's talk at the ISBA 2008 World meeting on Hamilton Island, we do not 
understand the purpose of the dismissal of MCMC methods found in the paper (``more for less", ``direct approximations", \&tc.) 
Convergence of MCMC methods has been the core activity of many top researchers in the past two decades, first and foremost 
Gareth Roberts and Jeff Rosenthal, whose work cannot be so casually ignored! Especially when considering that, first, the 
main appeal in using particle methods \citep{gordon:salmon:smith:1993} is in handling massive data flux at 
frequencies MCMC cannot face---and this stands quite separate from a convergence issue---and, secondly, the 
body of work produced by the authors as listed in the reference list does not include any in-depth study of 
the convergence properties of the PL method. 

As also argued in other discussions therein,
the lack of warning in \cite{lopes:carvalho:johannes:polson:2010} or in previous papers about the unavoidable degeneracy of
the method is more than puzzling, as the authors undoubtedly are aware of this.  The short paragraph about Monte Carlo errors
contained in the current paper can be construed to be misleading in this regard since the Monte Carlo error 
$C_t/\sqrt{n}$ does not account for the resampling step. As demonstrated in the discussion by Robert and Ryder, the error 
may end up being $\text{O}(1/t)$ and miss the standard $\sqrt{n}$ Monte Carlo convergence. The corpus of work thus produced so far
seems to limit itself to the PL processing of an increasing sequence of state-space and dynamic examples where the PL method does
produce a reasonable output, but this series of case-studies does not constitute a sufficient validation in our eyes. (Some of 
the examples processed in the current paper are missing the hyperparameters chosen for their satisfactory resolution.)

We note as a side remark that the argument found in Section 1.5 of the current paper about the difficulty about the improper prior $p(\theta)$ 
being solved by using the mixture representation 
$$
p(\theta) = \int p(\theta|Z_0)\,p(Z_0)\ted Z_0
$$
is nonsensical as currently stated: if the marginal distribution of $\theta$ is improper, so is the joint distribution. 
We also fail to understand where in the paper the authors manage to take a ``new look at Bayes's theorem". If by this
they mean the decomposition used in the first page of Section 1, 
this is a standard hidden Markov model property \citep{cappe:moulines:ryden:2004}.

\locsection{Particle learning on mixtures}
In the case of a mixture of $k$ Poisson distributions,
$$
f(x|\omega,\mu) = \sum_{i=1}^k p_i g(x|\lambda_i)\,,
$$
taken as a first example in \cite{carvalho:lopes:polson:taddy:2009}, the integrated predictive 
$p(y_{t+1}|\ev_t)$ can be obtained in closed form, as detailed below (since this derivation is 
central to our own Monte Carlo experiment). For Poisson mixtures, the ``essential" auxiliary variable
is $\ev_t=(n_1^t,\ldots,n_k^t,s_1^t,\ldots,s_k^t)$, where $n_i^t$ denotes the number of observations allocated
to the $i$-th component and $s_i^t$ the sum of the observations allocated to component $i$ $(1\le i\le k)$.

Thus, under a conjugate Dirichlet-Gamma prior assumption, using the delta function $\delta_{i=j}$ and the notation
$\gamma_\cdot=\gamma_1+\cdots+\gamma_k$,
\begin{align*}
p(y_{t+1}|\ev_t) 
&= \int p(y_{t+1}|\theta)p(\theta|\ev_t)\ted\theta\\
&= \int \left\{ \sum_{i=1}^k p_i \dfrac{\lambda_i^{y_{t+1}} e^{-\lambda_i}}{y_{t+1}!}\right\}
\, \mathcal{D}(p|(\gamma_j+n_j^t))\prod_{j=1} \mathcal{G}(\lambda_j|\alpha_j+s_j^t,\beta_j+n_j^t)\,\ted p\,\ted\lambda\\
&= \int \sum_{i=1}^k \dfrac{\Gamma(\gamma_\cdot+t)}{\Gamma(\gamma_i+n_i^t)}\,
    \dfrac{\Gamma(\gamma_i+n_i^t+1)}{y-{t+1}!\Gamma(\gamma_\cdot+t+1)}\,
    \dfrac{\Gamma(\alpha_i+s_i^t+y_{t+1})}{\Gamma(\alpha_i+s_i^t)}
   \,\dfrac{(\beta_i+n_i^t)^{\alpha_i+s_i^t}}{(\beta_i+n_i^t+1)^{\alpha_i+s_i^t+y_{t+1}}}\\
&\quad\times \mathcal{D}(p|(\gamma_j+n_j^t+\delta_{i=j}))
\prod_{j=1}^k \mathcal{G}(\lambda_j|\alpha_j+s_j^t+\delta_{i=j}y_{t+1},\beta_j+n_j^t+\delta_{i=j})\,\ted p\,\ted \lambda\\
&= \sum_{i=1}^k \dfrac{\Gamma(\gamma_\cdot+t)}{\Gamma(\gamma_i+n_i^t)}\,
    \dfrac{\Gamma(\gamma_i+n_i^t+1)}{y-{t+1}!\Gamma(\gamma_\cdot+t+1)}\,
    \dfrac{\Gamma(\alpha_i+s_i^t+y_{t+1})}{\Gamma(\alpha_i+s_i^t)}
    \dfrac{(\beta_i+n_i^t)^{\alpha_i+s_i^t}}{(\beta_i+n_i^t+1)^{\alpha_i+s_i^t+y_{t+1}}}\\
&=  \sum_{i=1}^k \dfrac{\gamma_i+n_i^t}{\gamma_\cdot+t+1}\,\dfrac{1}{y_{t+1}!}
    \dfrac{\Gamma(\alpha_i+s_i^t+y_{t+1})}{\Gamma(\alpha_i+s_i^t)}\,
    \dfrac{(\beta_i+n_i^t)^{\alpha_i+s_i^t}}{(\beta_i+n_i^t+1)^{\alpha_i+s_i^t+y_{t+1}}}
\end{align*}
gives a closed form of the predictive distribution of $y_{t+1}$ given the sufficient (simulated) statistic $\ev_t$. 

Furthermore, the distribution of the first version of the auxiliary variable, $z_1$, is easily derived, as
\begin{align*}
\BP(z_1=j|y_1) &\propto \int p_j g(y_1|\lambda_j) \,\pi(p,\lambda)\,\ted p\,\ted \lambda\\
&\propto \int \lambda_j^{y_{1}+\alpha_j-1} e^{-\lambda_j(1+\beta_j)} \,p_j^{1+\gamma_j-1}
\,(1-p_j)^{1+\gamma_{\cdot}-\gamma_j-1}\,\ted p_j\,\ted \lambda_j\\
&= \Gamma(y_1+\alpha_j)\,(1+\beta_j)^{-(y_1+\alpha_j)}\,{\Gamma(1+\gamma_j)\Gamma(1+\gamma_{\cdot}-\gamma_j)}\big/{\Gamma(2+\gamma_{\cdot})}
\end{align*}
while the next allocations can be found as
\begin{align*}
\BP(z_{t+1}=j|y_{t+1},\ev_t) &\propto p(y_{t+1},k_{t+1}=j|\ev_t) 
= \int p(y_{t+1},k_{t+1}=j|\theta) \,\pi(\theta|\ev_t)\,\ted \theta\\
&\propto \dfrac{(\gamma_j+n_j^t)\Gamma(\alpha_i+s_i^t+y_{t+1})}{\Gamma(\alpha_i+s_i^t)}\,
\dfrac{(\beta_i+n_i^t)^{\alpha_i+s_i^t}}{(\beta_i+n_i^t+1)^{\alpha_i+s_i^t+y_{t+1}}}
\end{align*}
and simulating the parameters of the Poisson model given $\ev_t$ is obvious, provided one uses a conjugate prior
\citep{diebolt:robert:1990a}.

\begin{figure}[p]
\begin{center}
 \includegraphics[height=0.3\textwidth]{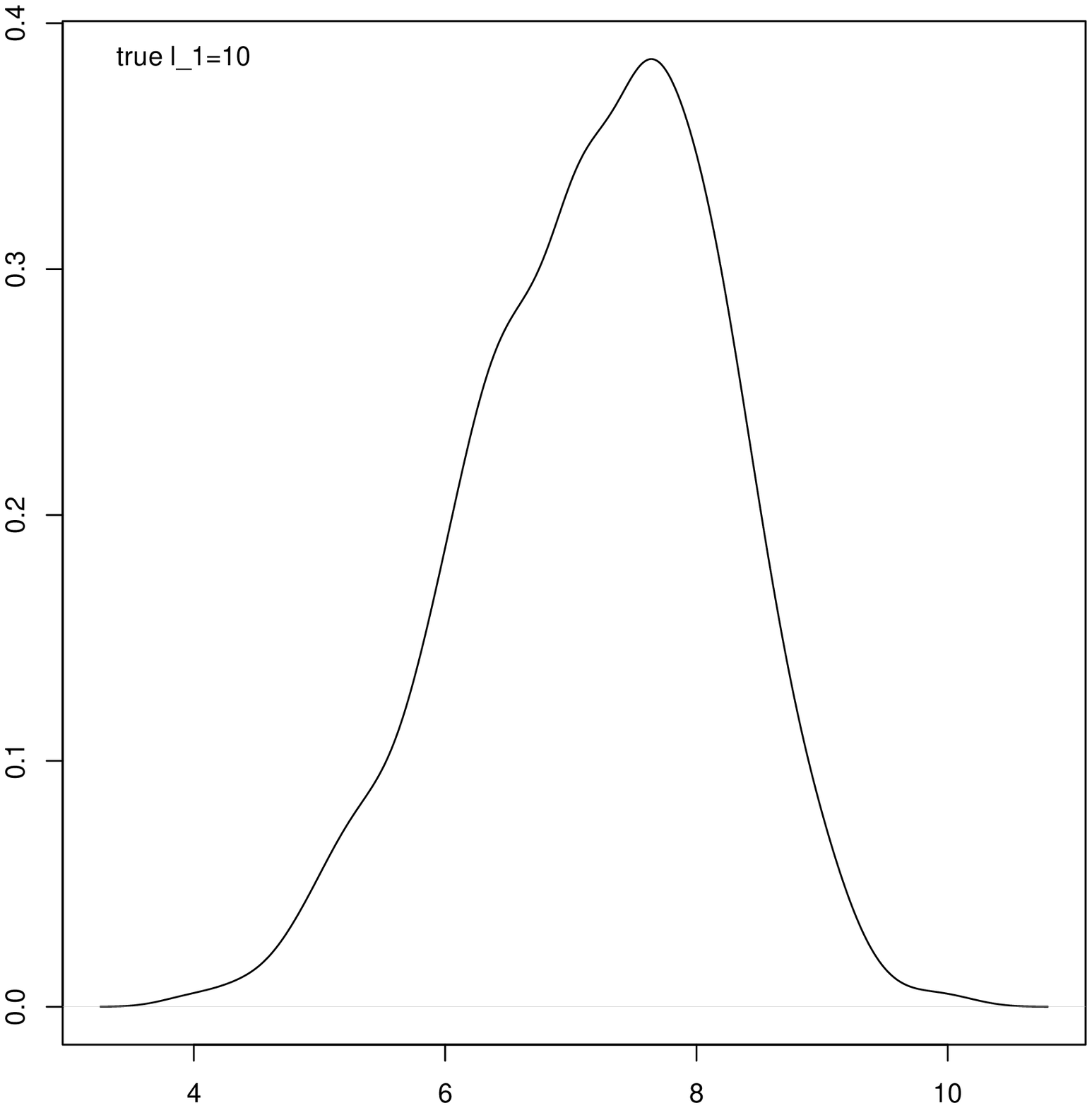}\includegraphics[height=0.3\textwidth]{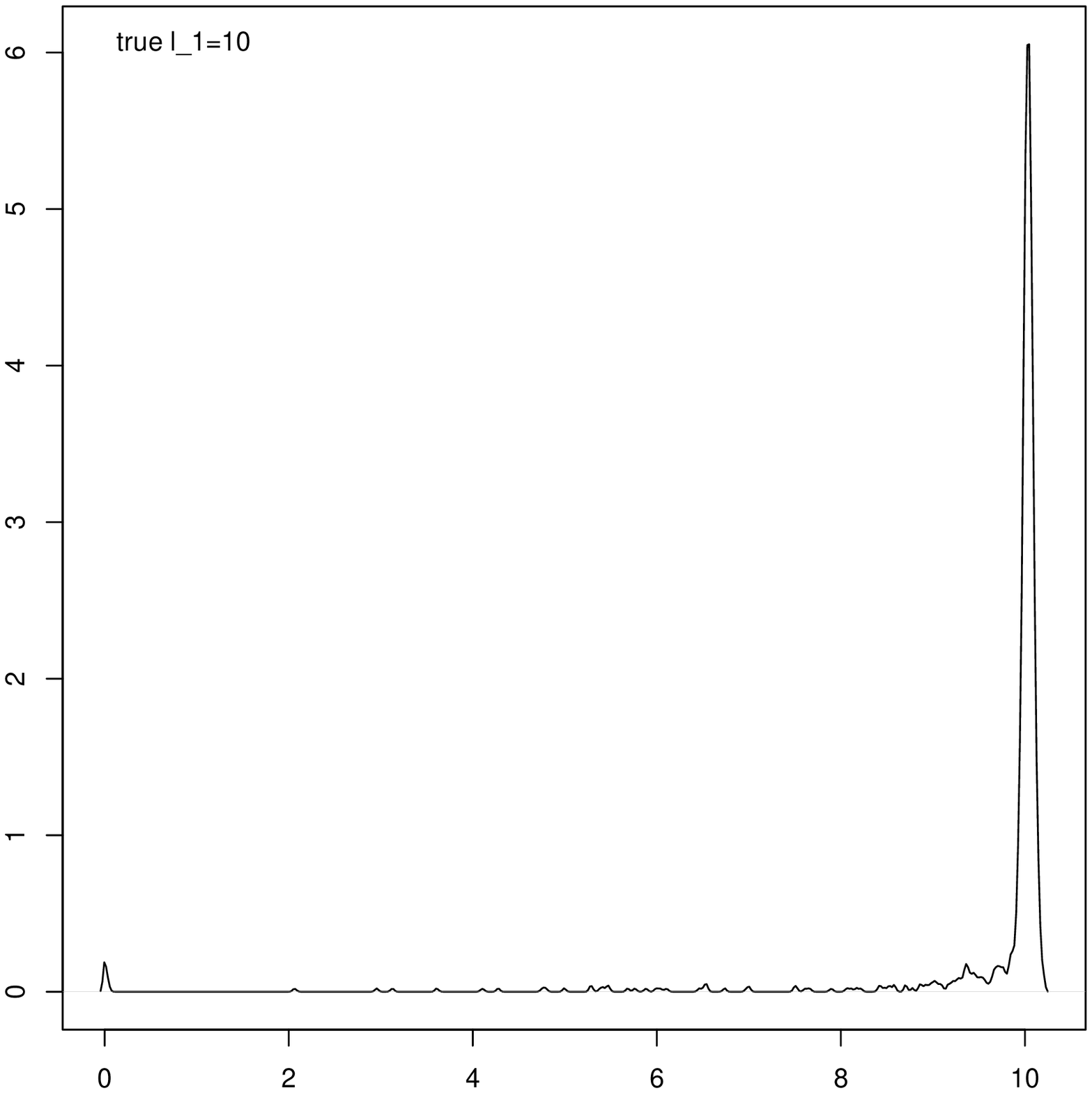}
 \includegraphics[height=0.3\textwidth]{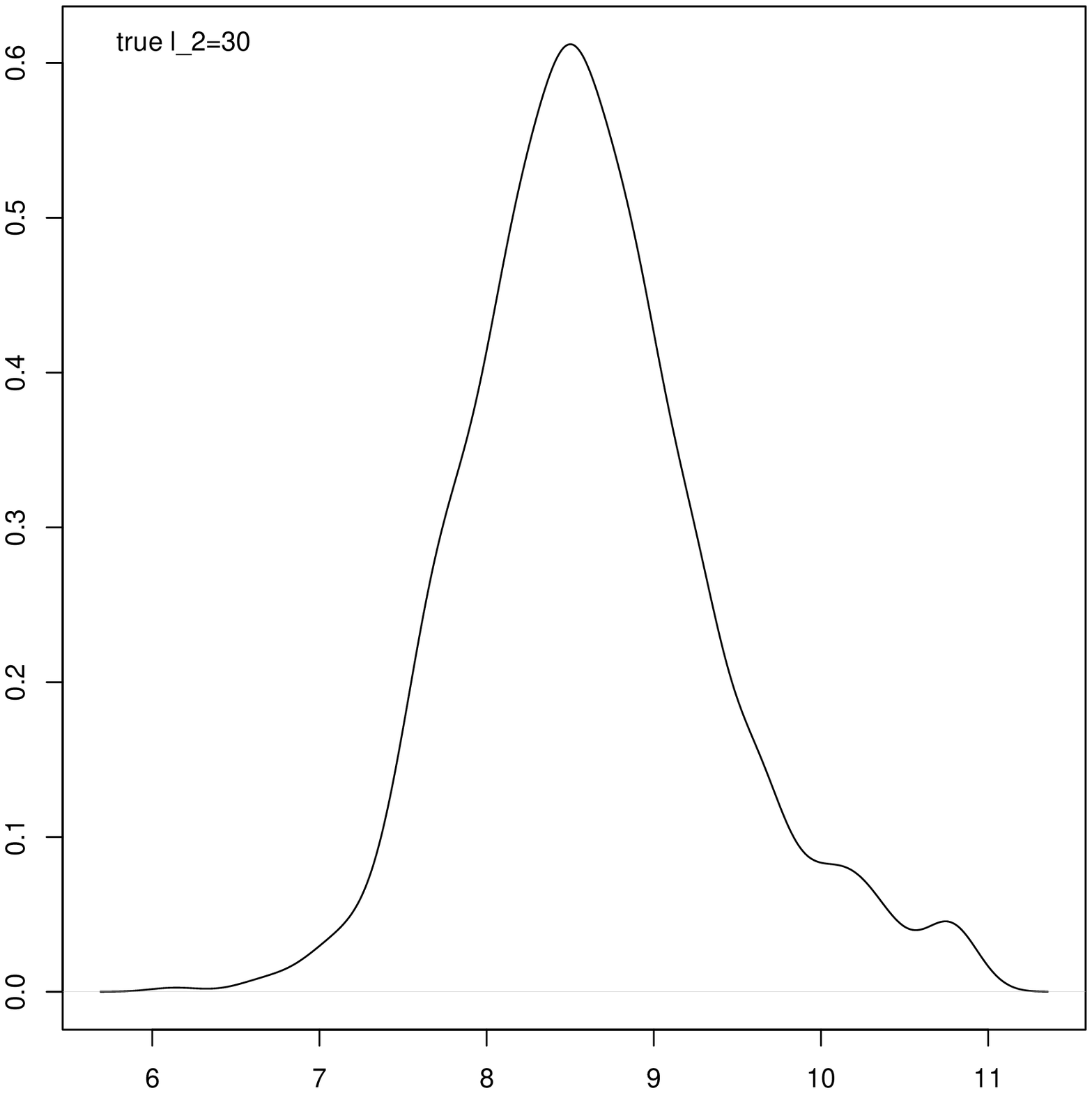}\includegraphics[height=0.3\textwidth]{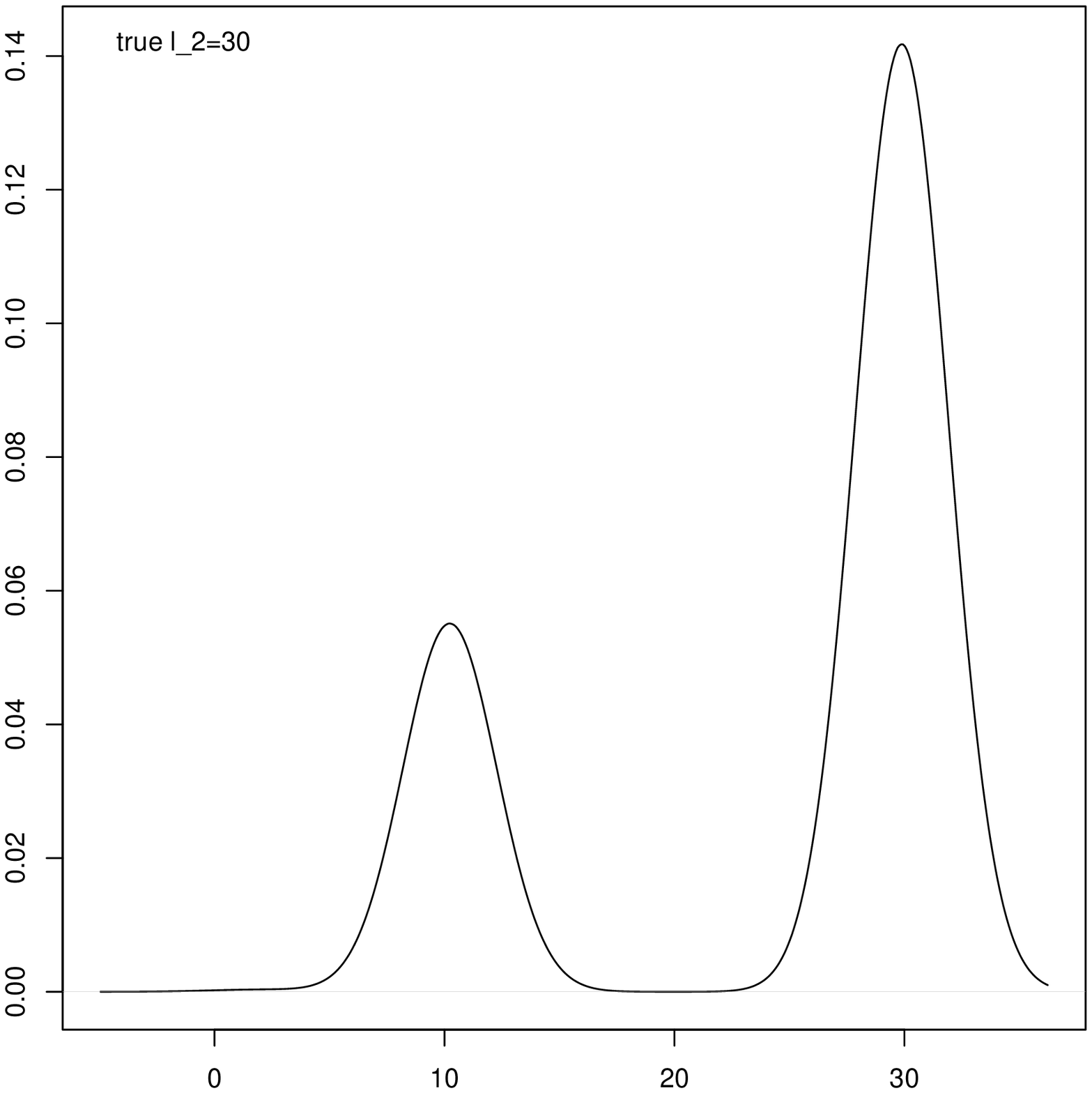}
 \includegraphics[height=0.3\textwidth]{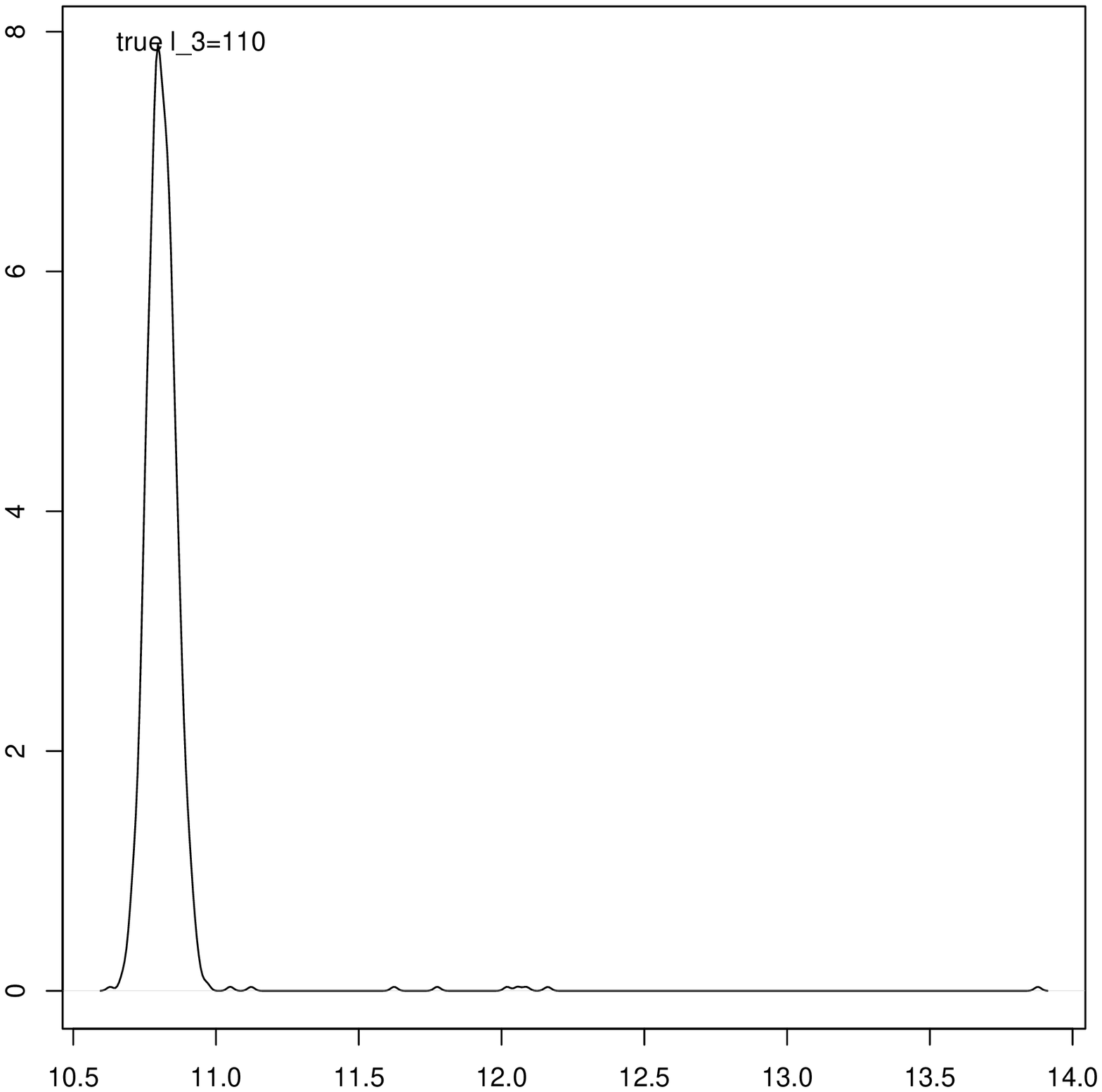}\includegraphics[height=0.3\textwidth]{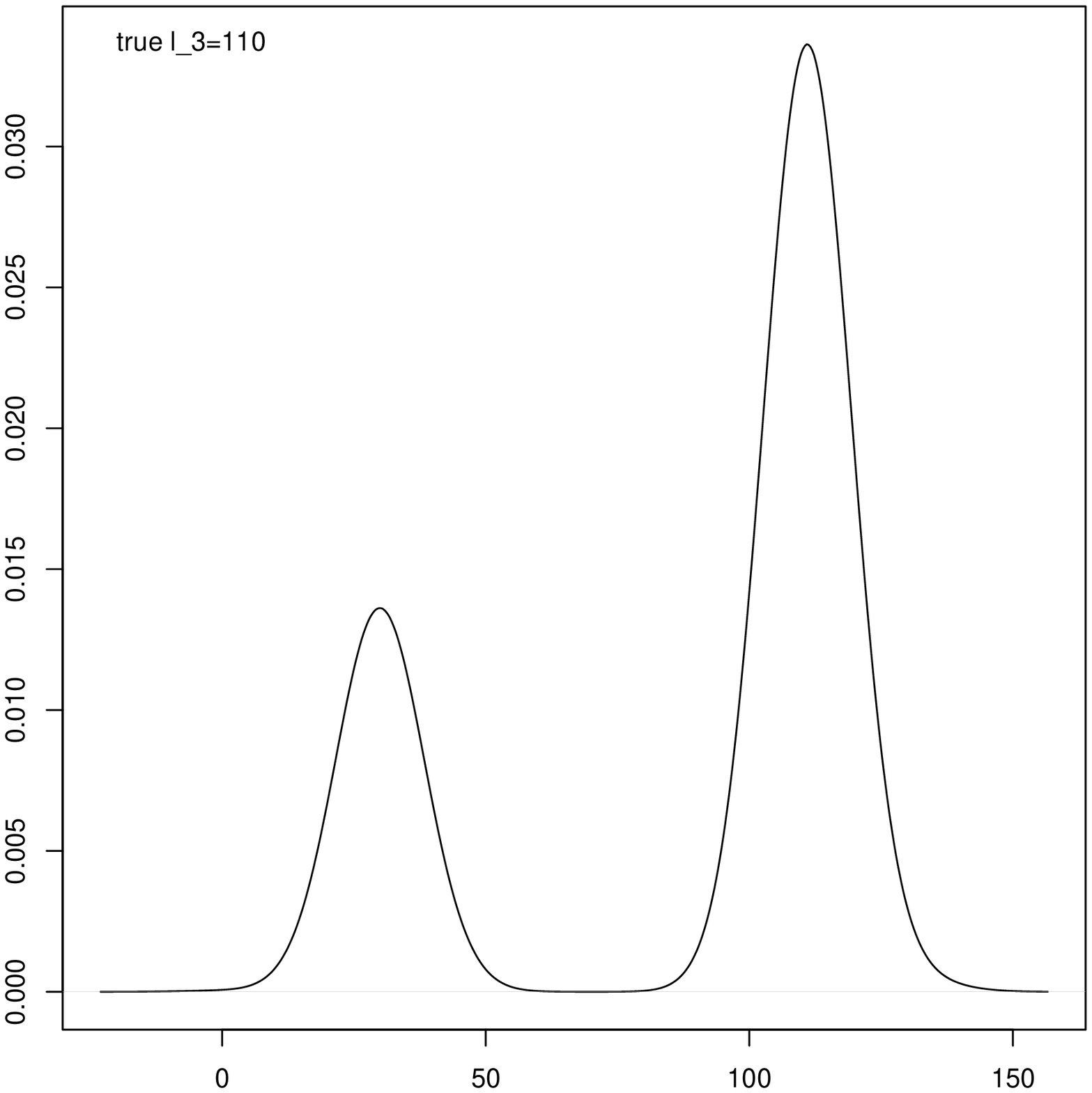}
 \includegraphics[height=0.3\textwidth]{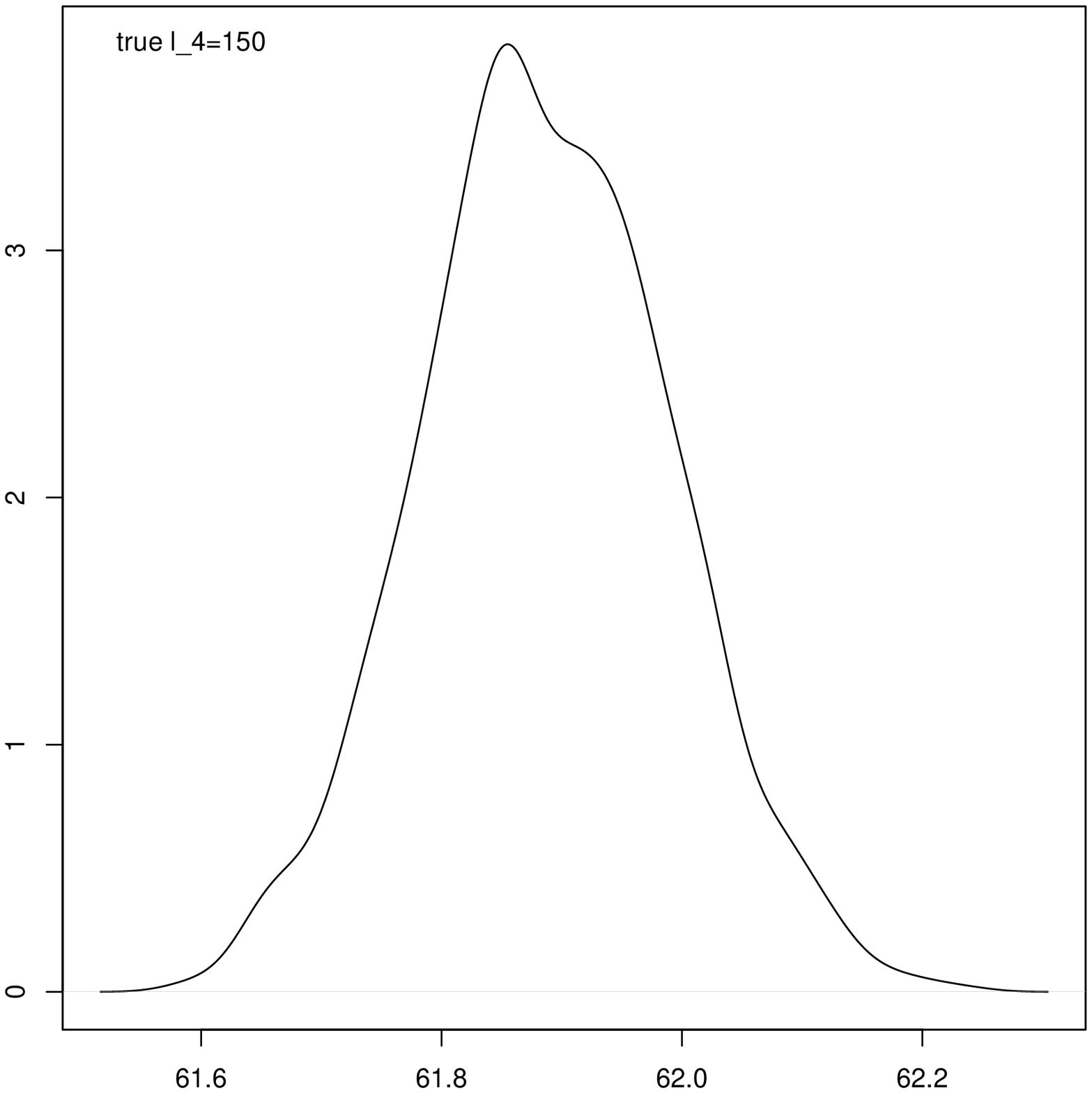}\includegraphics[height=0.3\textwidth]{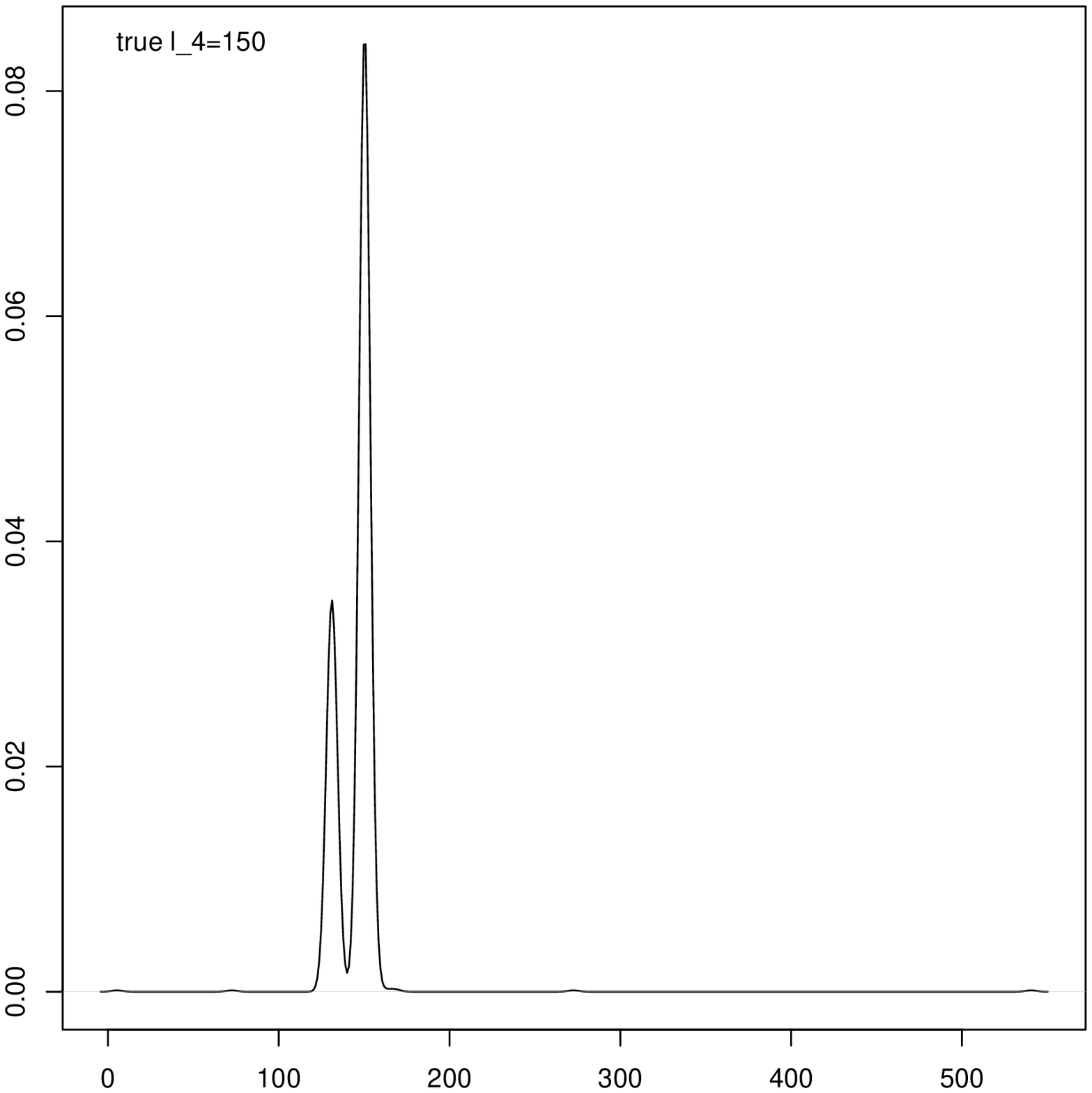}
\end{center}
\caption{\label{fig:compapos}
Comparison of the posterior distributions on the ordered $\lambda_i$'s produced by PL {\em (left)} and MCMC {\em (right)} for $10^4$ 
simulated observations from a $4$ component Poisson mixture and $10^4$ particles/MCMC iterations. The curves are
obtained by apply the {\sf R} function {\sf density()} to both samples. The true values are indicated at the top
of each graph}
\end{figure}

When applying both PL and MCMC techniques to a sample of size $10^4$ extracted from the Monte Carlo study detailed in the
discussion by Iacobucci et al., we obtain the output represented in Figure \ref{fig:compapos} for the posterior
distributions of the $\lambda_i$'s. (Both PL and MCMC samples were re-ordered in terms of the $\lambda_i$'s. The
plots are therefore the posterior distributions of the order statistics $\lambda_{(i)}$.) The discrepancy between
both approaches is clear on this example. (We stress that this represents a ``worst case" in the
sense that the observations were chosen from the above-mentioned Monte Carlo experiment 
by selecting the sample producing the largest discrepancy in the evidence approximations. Random picks of samples from 
the Poisson mixture usually produce a better agreement.)

\section{On the approximation of evidence (Iacobucci, Robert, Marin and Mengersen)}\label{sec:vid}
\bigskip
In this discussion, we consider the performances of the particle learning (PL) technique 
in the specific setting of mixtures of distributions and for the approximation of the {\em evidence}
$$ 
\ev_i = \int_{\Theta_i} \pi_i(\theta_i) f_i(y|\theta_i)\,\text{d}\theta_i\,,
$$
aka the marginal likelihood. Through a simulation experiment, we examine how much the degeneracy 
that is inherent to particle systems impacts this approximation (We refer the reader to 
\citealp{chen:shao:ibrahim:2000}, for a general approach to the approximation of evidence 
and to both \citealp{chopin:robert:2010}, and \citealp{marin:robert:2010}, for illustrations 
in the particular setting of mixtures.) 

\locsection{Approximation of the evidence}
In the case of a mixture of $k$ Poisson distributions,
$$
f(x|\omega,\mu) = \sum_{i=1}^k p_i g(x|\lambda_i)\,,
$$
taken as an example in \cite{lopes:carvalho:johannes:polson:2010},  
and studied in \cite{carvalho:lopes:polson:taddy:2009}
the integrated predictive can be obtained in closed form, as derived in the discussion of Mengersen
et al. This implies that the product approximation to the evidence
$$
p(y^t) = \prod_{r=1}^t p(y_r|y^{r-1}) \approx \prod_{r=1}^t \frac{1}{N}\sum_{i=1}^N
p(y_r|\ev_{r-1}^{(i)})
$$
proposed in \cite{carvalho:lopes:polson:taddy:2009} and \cite{lopes:carvalho:johannes:polson:2010} can be implemented 
here. We thus use the setting of Poisson mixtures to evaluate this PL approximation of the evidence and we re-evaluate 
Carvalho et al.'s (\citeyear{carvalho:lopes:polson:taddy:2009})
assessment that this {\em ``approach offers a simple and robust sequential Monte Carlo alternative
to the traditionally hard problem of approximating marginal predictive densities via MCMC output"}.

We note that, since the PL sample is considered as an approximate sample from the posterior 
$\pi(p,\lambda|y^t)$ it is possible to evaluate the evidence using \citeauthor{chib:1995}'s 
(1995) formula rather than the above proposal of the authors. The availability of an alternative 
estimator of the evidence allows for a differenciation between the evaluation of approximation 
[of the target posterior distribution] resulting from the particle system (seen through a possible 
bias in Chib's, 1995, version) and the evaluation of the approximation [of the evidence] resulting from 
the use of the product marginal in \cite{lopes:carvalho:johannes:polson:2010}. Thus, in contrast to
the other discussions of ours, we evaluate here the specific degeneracy of the evidence approximation
due to using a product of approximations.

\begin{figure}
\begin{center}
 \includegraphics[height=0.3\textwidth]{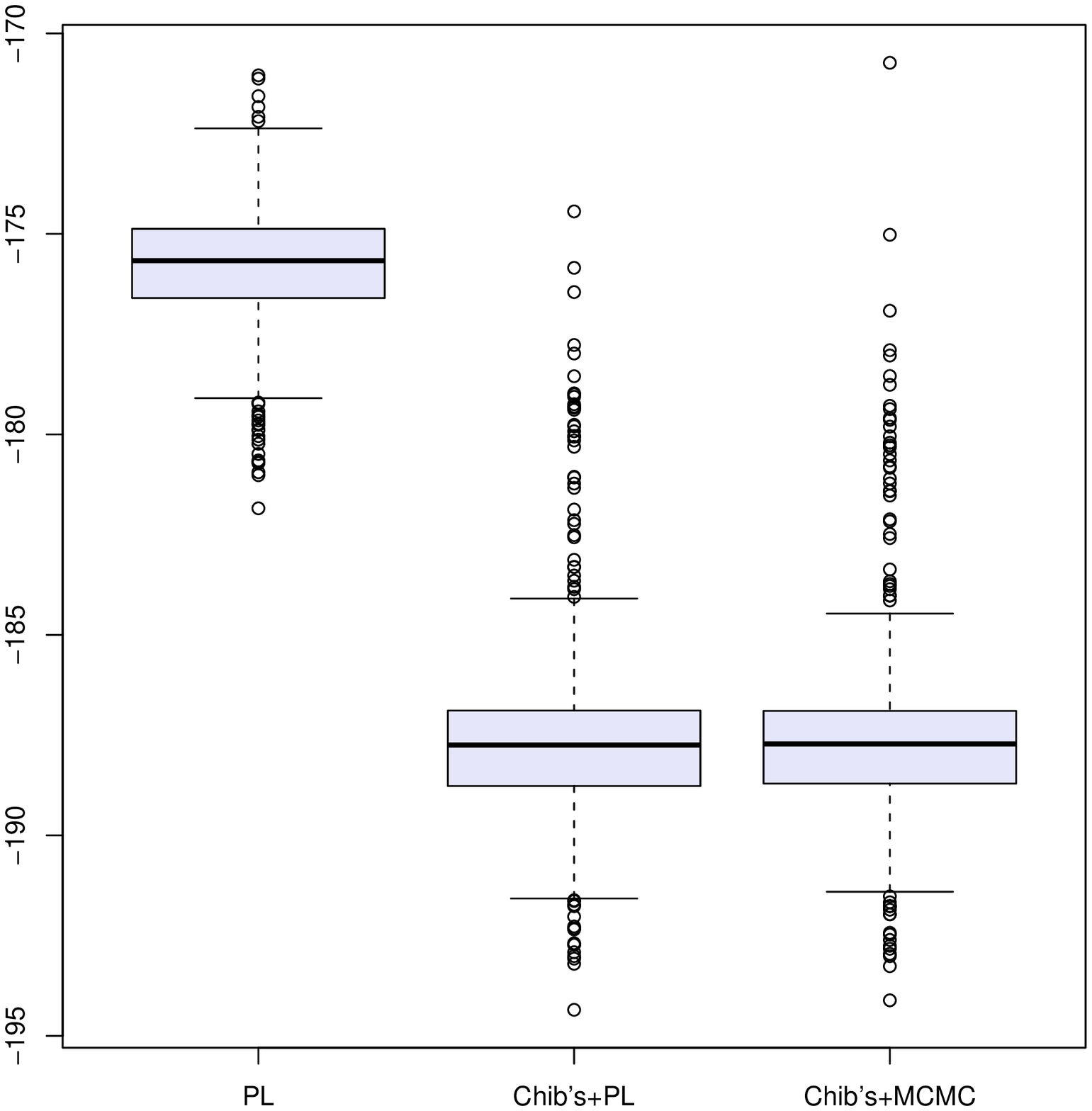}
 \includegraphics[height=0.3\textwidth]{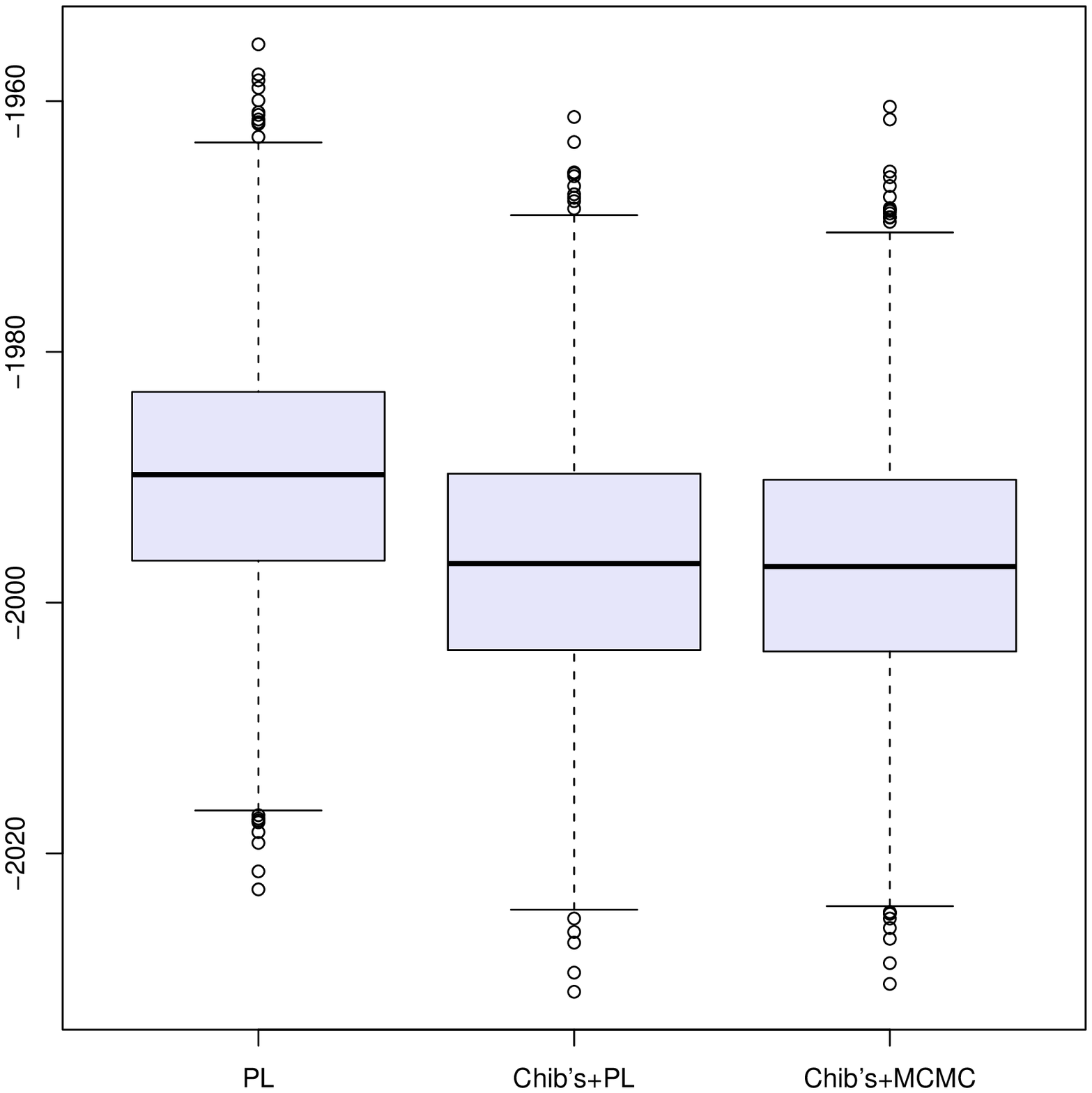}
 \includegraphics[height=0.3\textwidth]{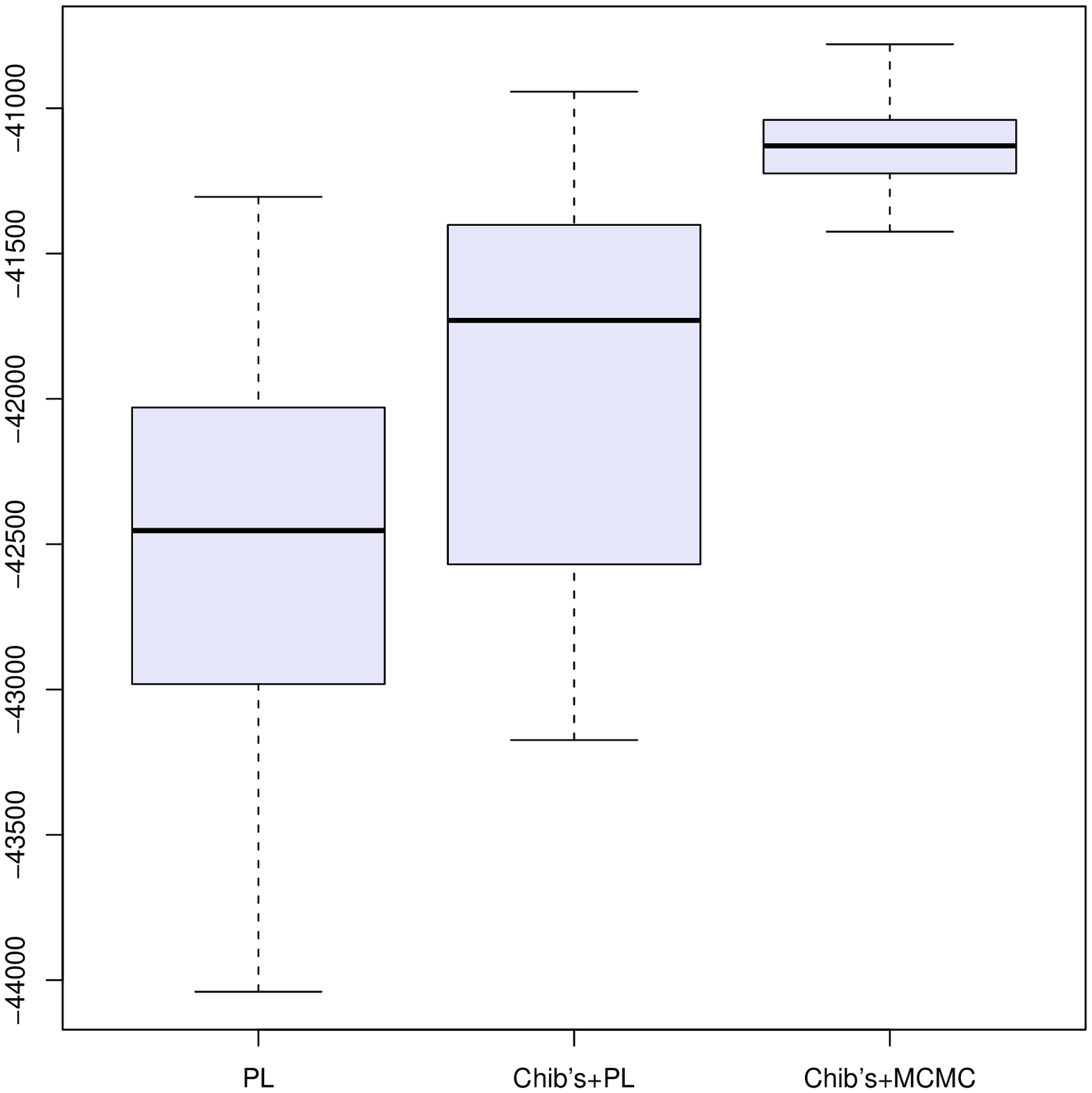}
\end{center}
\caption{\label{fig:kompa1}
Evolution against the number of observations ($n=100$, $n=1,000$ and $n=10,000$)
of the evidence approximation based on a PL sample and \cite{lopes:carvalho:johannes:polson:2010} approximation,
on a PL sample and Chib's (1995) approximation, on an MCMC sample and Chib's (1995) approximation,
for a particle population of size $10,000$, a mixture with $4$ components and 
scale parameters $\lambda=(10,50,110,150)$.}
\end{figure}

\begin{figure}
\begin{center}
 \includegraphics[height=0.3\textwidth]{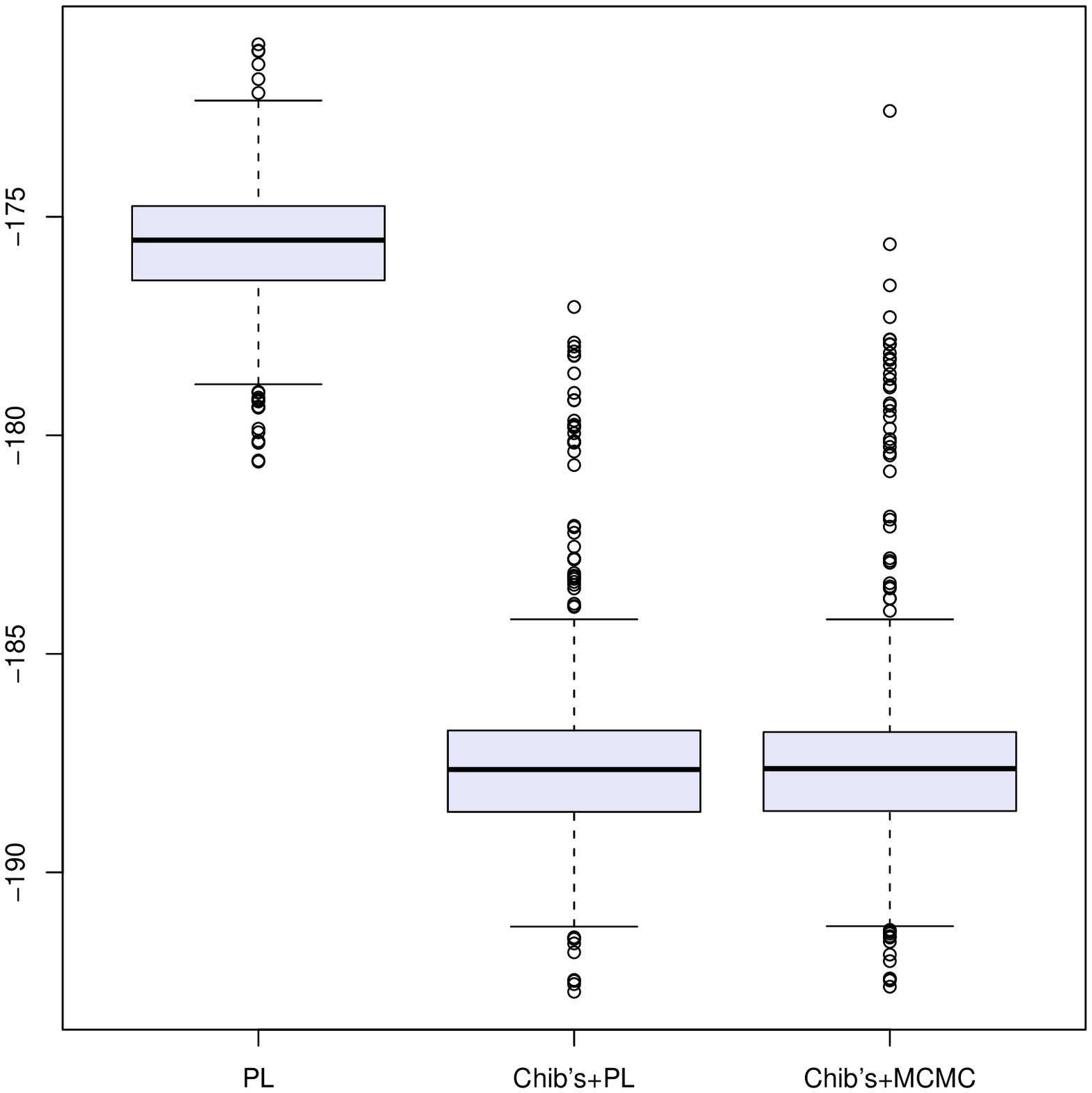}
 \includegraphics[height=0.3\textwidth]{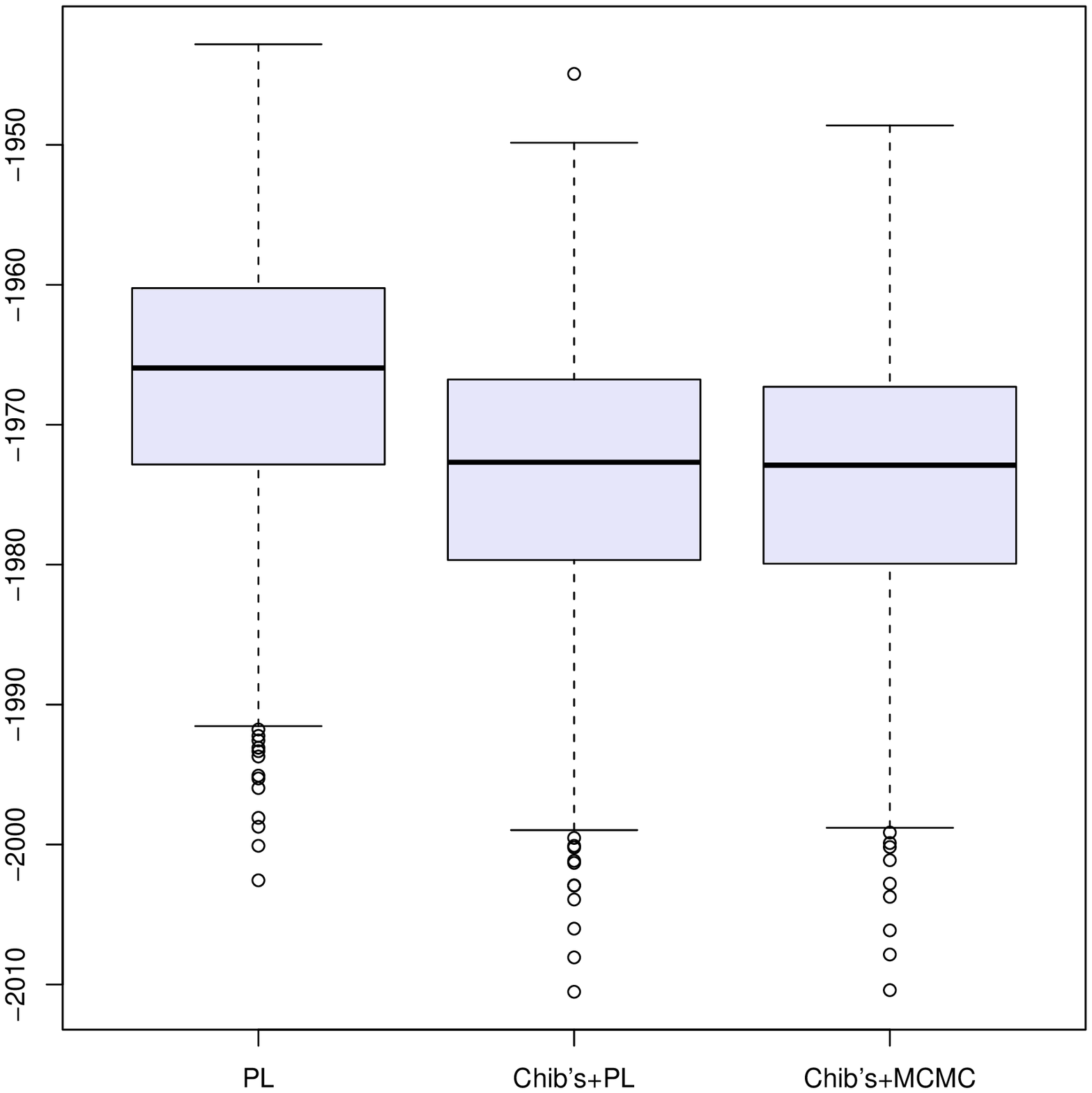}
 \includegraphics[height=0.3\textwidth]{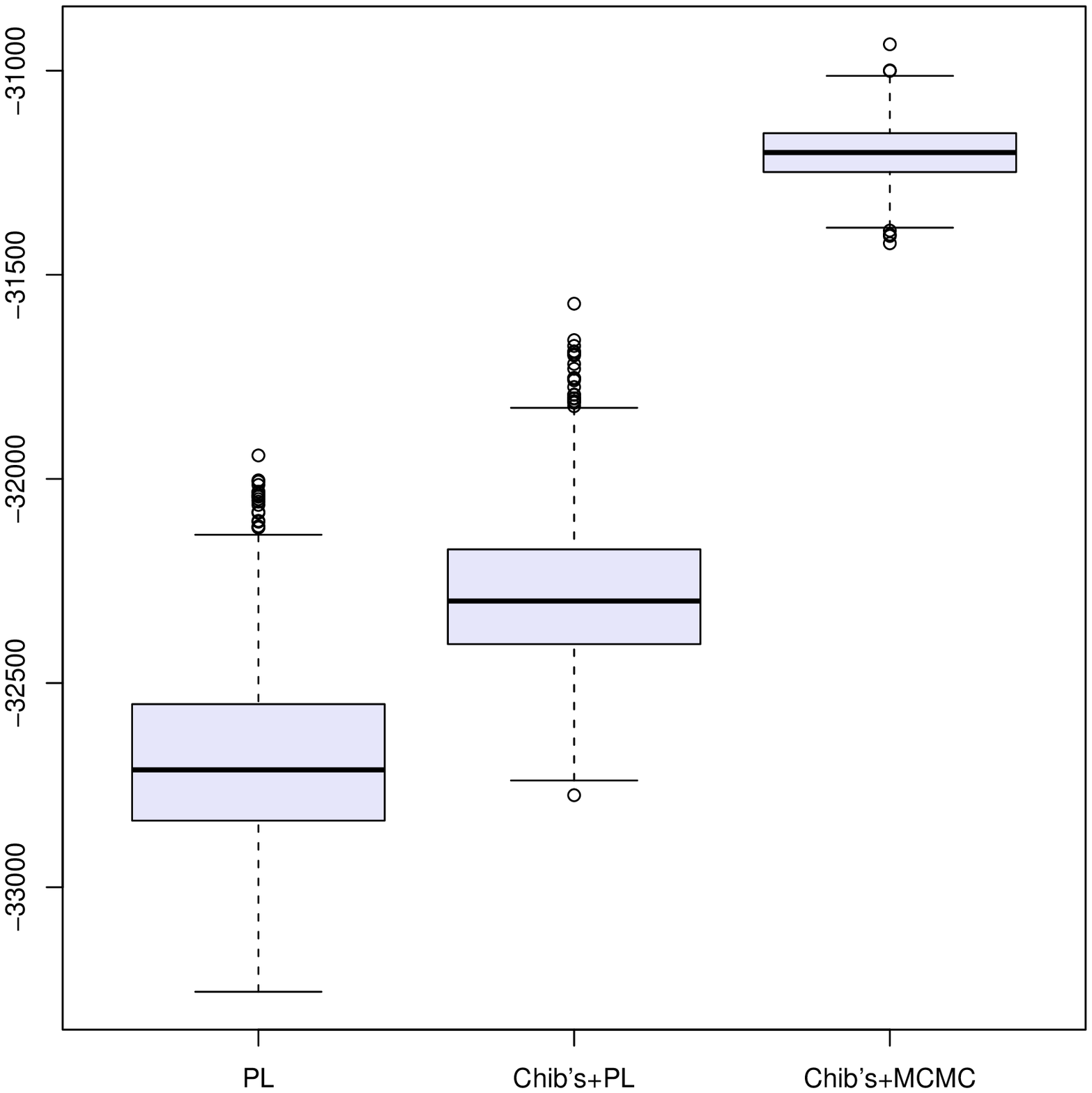}
\end{center}
\caption{\label{fig:kompa2}
Same caption as Figure \ref{fig:kompa1} for $\lambda=(10,15,20,25)$.}
\end{figure}

\begin{figure}
\begin{center}
 \includegraphics[height=0.3\textwidth]{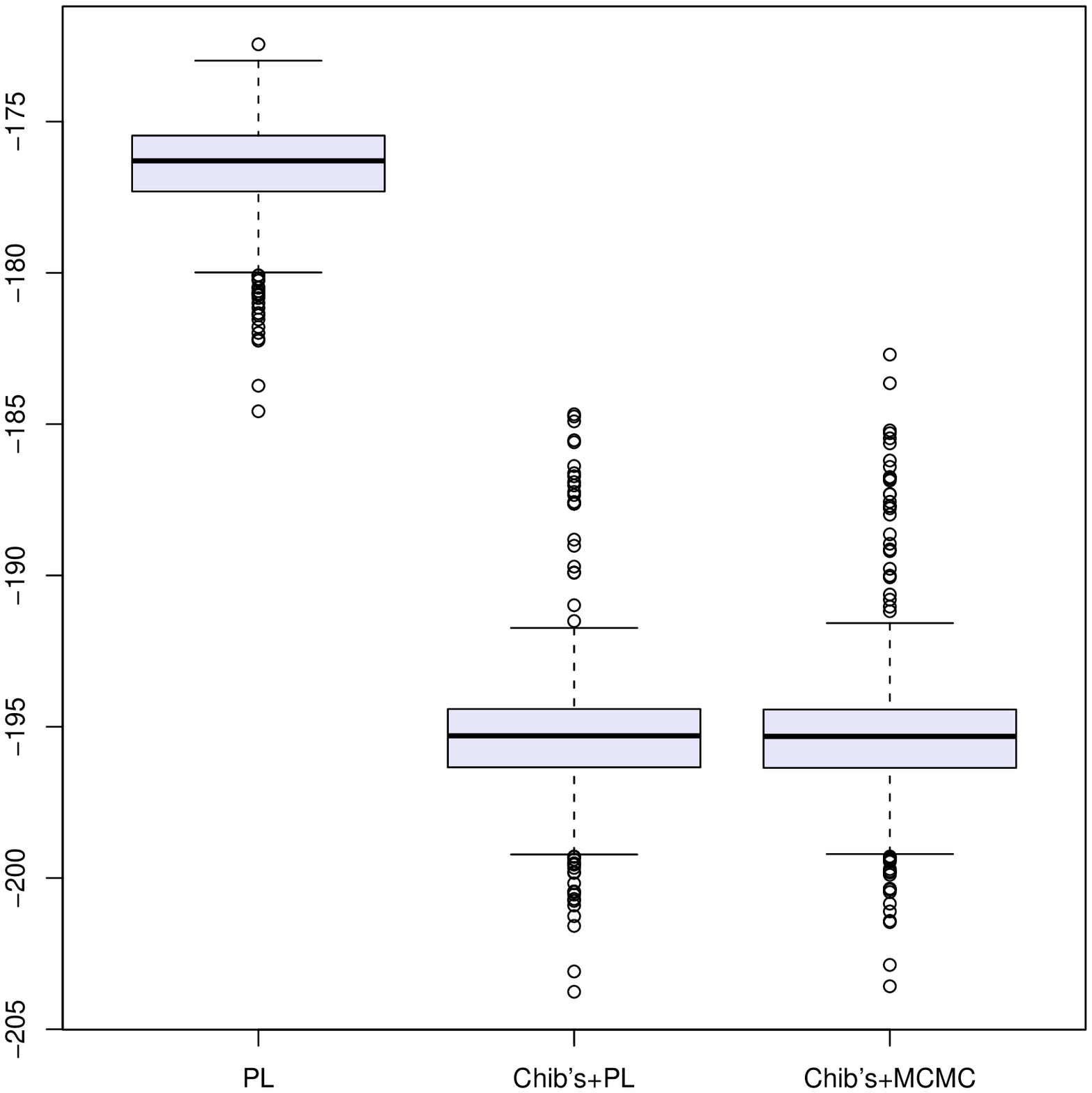}
 \includegraphics[height=0.3\textwidth]{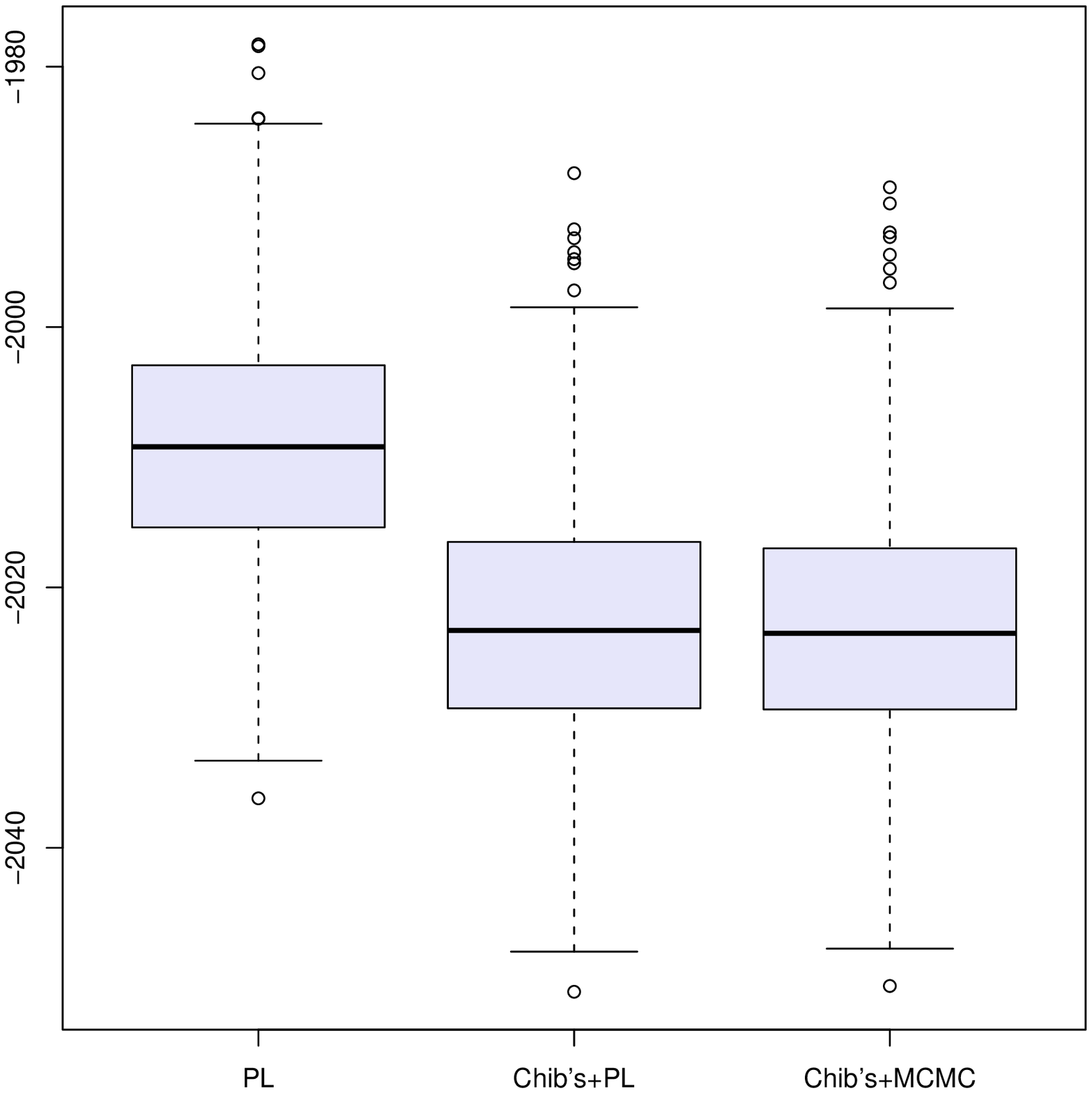}
 \includegraphics[height=0.3\textwidth]{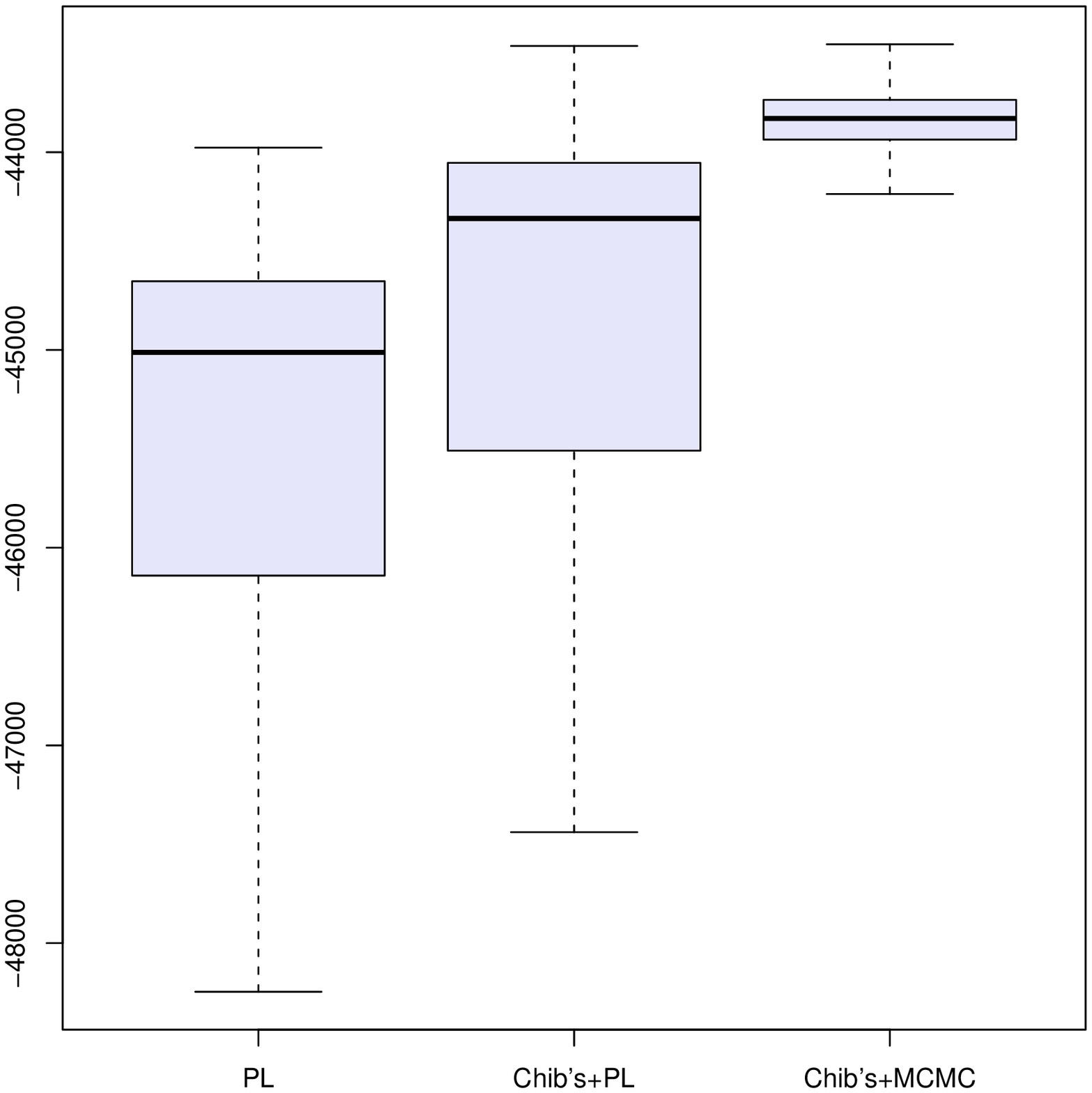}
\end{center}
\caption{\label{fig:kompa3}
Same caption as Figure \ref{fig:kompa1} for $5$ components.}
\end{figure}

\begin{figure}
\begin{center}
 \includegraphics[height=0.3\textwidth]{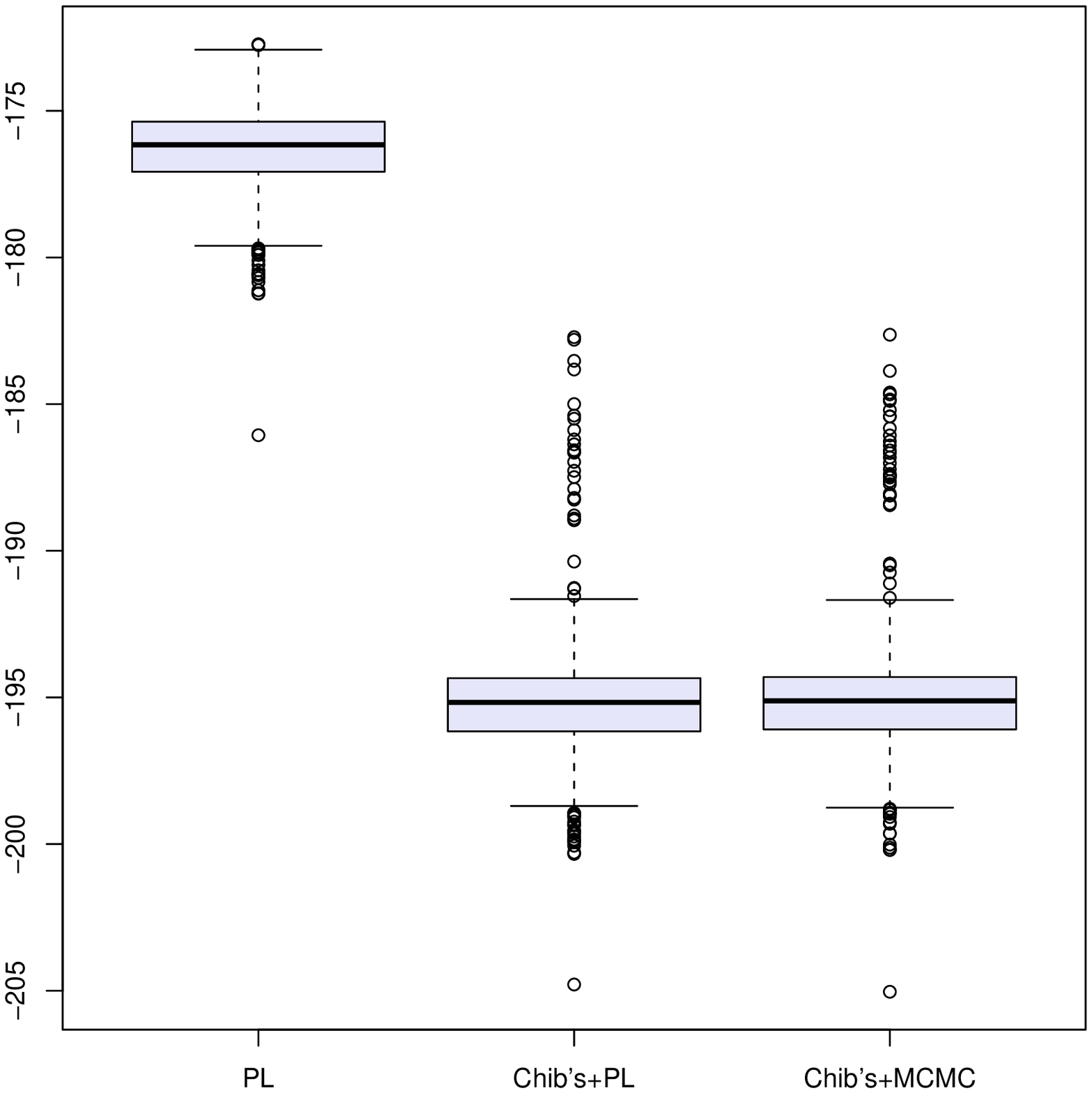}
 \includegraphics[height=0.3\textwidth]{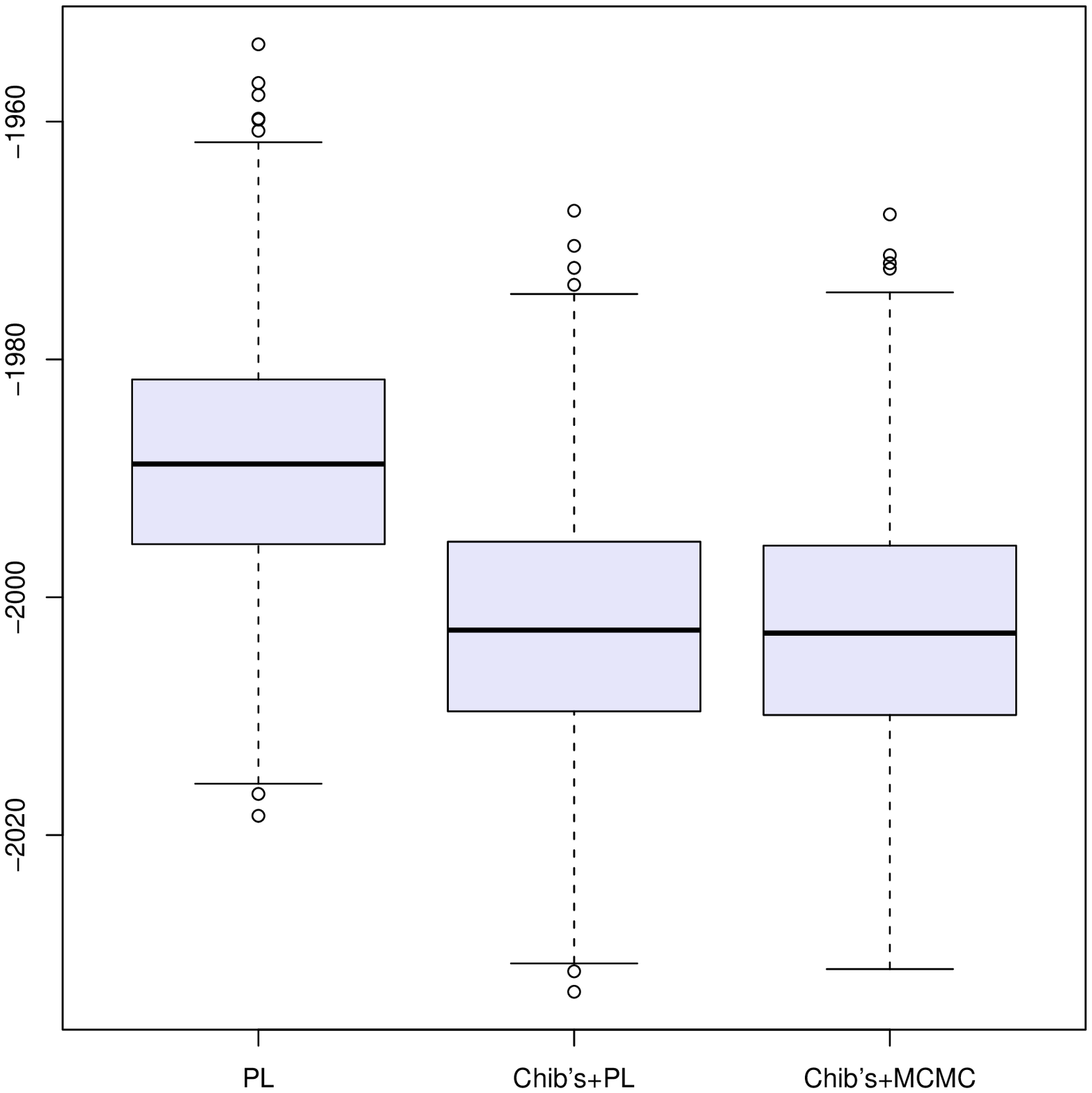}
 \includegraphics[height=0.3\textwidth]{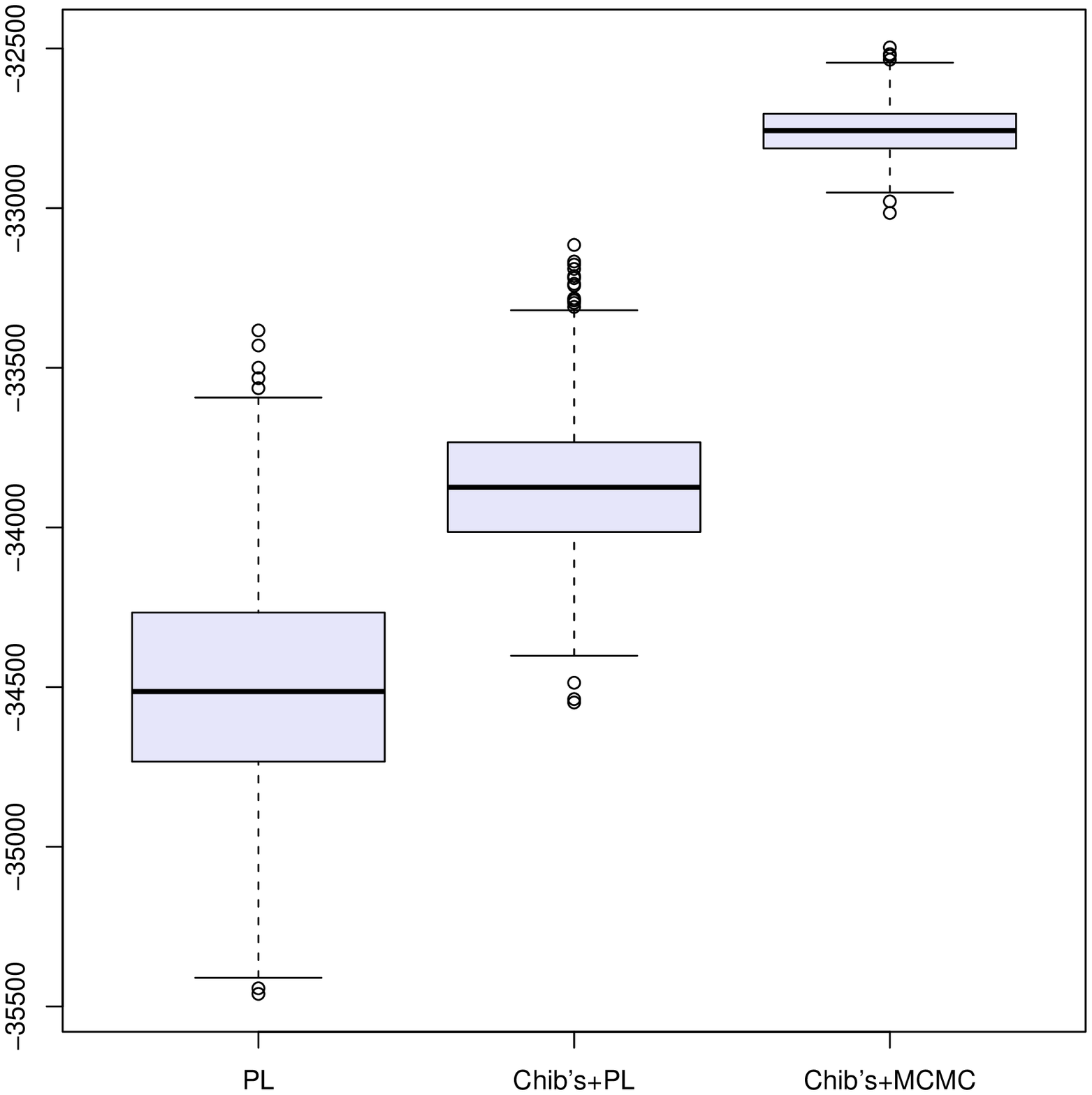}
\end{center}
\caption{\label{fig:kompa4}
Same caption as Figure \ref{fig:kompa2} for $5$ components.}
\end{figure}

\locsection{A Monte Carlo experimentation}
In order to evaluate the performances of the PL algorithm when compared with the vanilla Gibbs sampler
\citep{diebolt:robert:1990a,diebolt:robert:1994}, we simulated 250 samples of size $10^4$ from Poisson mixtures with
4 and 5 components and with either widely spaced or close components, $\lambda=(10,50,110,150,180,210)$ and
$\lambda=(10,15,20,25,30,35)$, respectively, and with slightly decreasing weights $p_i$. We ran a $10^4$ iteration
Gibbs sampler for Figures \ref{fig:kompa1}--\ref{fig:kompa4}, performing a further $10^6$ iterations 
as a check of the stability of
the MCMC approximation. (For Chib's approximation to perform correctly, as noted in \citealp{berkhof:mechelen:gelman:2003} and
\citealp{marin:robert:2010}, it is necessary to average over all $k!$ permutations of the component indices
for both the original PL sample and the MCMC sample in order to escape label switching issues.) 

The first interesting outcome of our experiment
is that the PL sample does not suffer from degeneracy for a small enough number of
observations, since the ranges of the \citeauthor{chib:1995}'s (2005) approximations for both 
PL and MCMC samples (represented by the second and third columns in the boxplots) are then the same.
However, as predicted by the theory (see the discussions by Chopin and Robert, and by Robert and Ryder), increasing the number 
of observations without simultaneously and exponentialy increasing the number of particles necessarily leads to the degeneracy 
of the simulated sufficient statistic paths. In our experiment, this degeneracy 
always occurs between $5,000$ and $10,000$ observations. 
The phenomenon clearly appears on Figures \ref{fig:kompa1}--\ref{fig:kompa4} where both the range 
and the extremes of the evidence approximations significantly differ on the right hand side boxplot graph. (Again, the
stability of the MCMC range was tested by running the Gibbs sampler for much longer and observing no variation.)
This divergence is to be contrasted with Figure 1 in \cite{carvalho:lopes:polson:taddy:2009} which concludes
to an agreement between all approximations to the Bayes factor.

\begin{figure}
\begin{center}
 \includegraphics[height=0.5\textwidth]{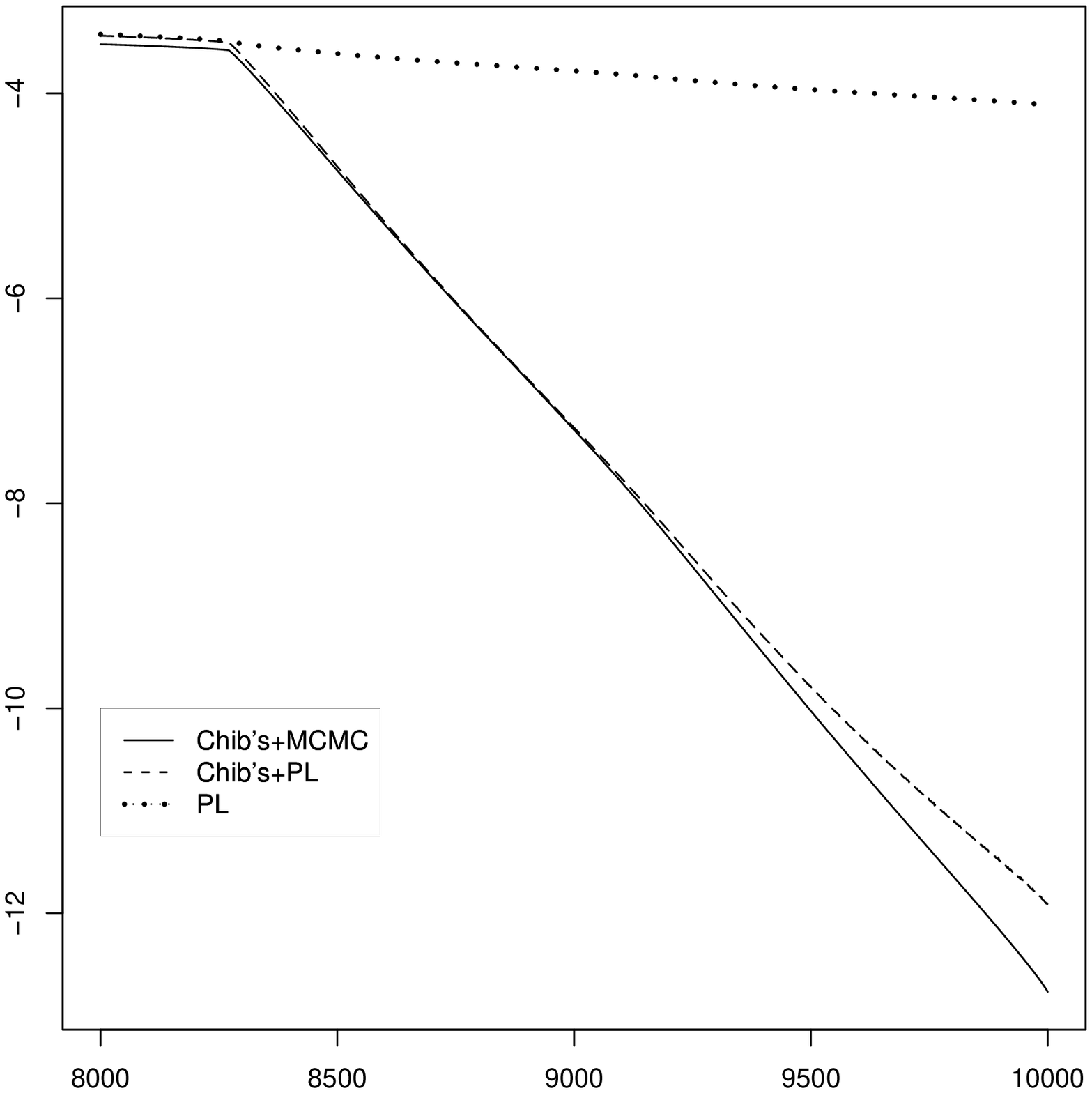}
\end{center}
\caption{\label{discrr}
Evolution of the three approximations of the evidence {\em per observation} against the number of 
observations for a specific sample simulated from the same Poisson mixture as in Figure \ref{fig:kompa1}.}
\end{figure}

The second result that is relevant for our discussion is that the new approximation to the evidence proposed by the authors
suffers from a severe bias as one proceeds through the observations. This issue is apparently unrelated to the degeneracy 
phenomenon observed above in that the discrepancy starts from the beginning, the closest approximation
occuring for $n=1,000$ observations. Note that \cite{carvalho:lopes:polson:taddy:2009}
mention that the evidence approximation based on particle learning was less variable. While this feature is not visible in our
experiment, it is not necessarily a positive feature in any case, as shown in the current experiment. 
(In order to provide a better rendering of the
comparison between the PL and the MCMC algorithms, we excluded the outliers from all boxplots. We however stress that 
both PL approaches had a higher propensity to outlying behaviour.) In the strongest case of discrepancy
between PL and MCMC found in our experiment, Figure \ref{discrr} illustrates the departure between the three approaches from
a particularly influential observation, since the graphs are compared in terms of evidence {\em per observation}.

We thus conclude at the lack of robustness of the new approximation of evidence suggested 
in both \cite{carvalho:lopes:polson:taddy:2009} and \cite{lopes:carvalho:johannes:polson:2010} 
(besides providing a reinforced demonstration of the overall difficulty with degeneracy).

\section{Repeatability of the degeneracy (Iacobucci, Marin and Robert)}
\begin{figure}
\begin{center}
 \includegraphics[height=0.4\textwidth]{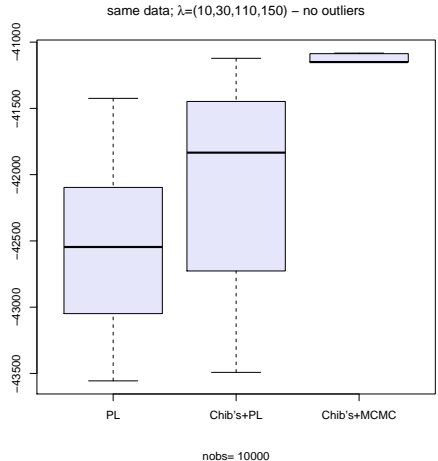}
\end{center}
\caption{\label{fig:kompa0}
Range of the evidence approximation based on a PL sample and \cite{lopes:carvalho:johannes:polson:2010} approximation,
on a PL sample and Chib's (1995) approximation, on an MCMC sample and Chib's (1995) approximation,
for a particle population of size $10,000$, a mixture with $4$ components and 
scale parameters $\lambda=(10,50,110,150)$, and 683 replications.}
\end{figure}

\bigskip
Following the floor discussion at the conference, we want to point out here
that the divergence between the evidence evaluations observed in
the discussion of Iacobucci et al.~is not the result of an outlying Monte Carlo experiment but indeed
a distributional property. This can be seen on Figure \ref{fig:kompa0}
which reproduces the study of Iacobucci et al. (in this set of comments) on the variation of the evidence
in the specific setting of mixtures of Poisson distributions. For two given datasets,
we repeated 683 times the three evidence approximations using the method proposed in Lopes et al. (2010),
and Chib's (1995) method applied to both the PL and MCMC samples. The divergence between the three evaluations
is consistent across simulations, so repeating simulations does not help in exhibiting this divergence.

\section{On degeneracy (Chopin and Robert)}
\bigskip
In this discussion, we consider the performance of the particle learning technique 
of \cite{lopes:carvalho:johannes:polson:2010} in a limiting case, in order to illustrate 
the fact that a particle system cannot but degenerate, even when considering sufficient statistics
$Z_t$ with fixed dimensions.

\locsection{Particle system degeneracy}
When \cite{lopes:carvalho:johannes:polson:2010} state that {\em $p(Z^t|y^t)$ is not
of interest as the filtered, low dimensional $p(Z_t|y^t)$ is sufficient for inference at time $t$}, 
they seem to implicitely imply that the restriction of the simulation focus to a low dimensional vector is a way to avoid
the degeneracy inherent to all particle filters (see, e.g., \citealp{delmoral:doucet:jasra:2006}).
However, the degeneracy of particle filters is an unavoidable consequence of the explosion of the state vector 
$Z^t$ and the issue does not vanish because one is only interested in the marginal
$$
p(Z_t|y^t) = \int p(Z^t|y^t) \,\ted Z^{-t}\,.
$$
Indeed, as shown by the pseudo-code rendering in \cite{lopes:carvalho:johannes:polson:2010}, the way PL produces
a sample from $p(Z_t|y^t)$ is by sequentially simulating $Z^t$ and by extracting $Z_t$ as the final output from this sequence.
The PL algorithm therefore relies on an approximation of $p(Z^t|y^t)$ and the fact that this approximation quikckly degenerates 
as $t$ increases, as discussed below and in the companion discussion by Robert and Ryder, 
obviously has an impact on the approximation of $p(Z_t|y^t)$.

Inherently, particle learning (PL) is at its core an auxiliary particle filter \citep{pitt:shephard:1999}
applied in settings where there exists a sufficient statistic \citep{darmois:1935} 
of reduced (or, even better, with fixed) dimension. The simulation scheme thus relies on resampling 
\citep{rubin:1988,kitagawa:1996} for adjusting the distribution of the current particle population 
to the new observation $y_{t+1}$. Because of this continual resampling, the number of different 
values of $Z_p$ $(p\ge 1)$ contributing to the sufficient statistic $Z_t$ $(t>p)$ is decreasing in $t$ at an exponential rate for 
a fixed $p$. Therefore, unless the size of the particle population exponentially increases with 
$t$ (see \citealp{douc:cappe:moulines:robert:2002}, and the companion discussion by Chopin and Sch\"afer), the sample of 
$Z_t$'s will not be distributed as an iid sample from $p(Z_t|y^t)$. The following section very clearly makes
this point through a simple if representative example.

\begin{figure}
\begin{center}
 \includegraphics[height=0.8\textwidth,angle=270]{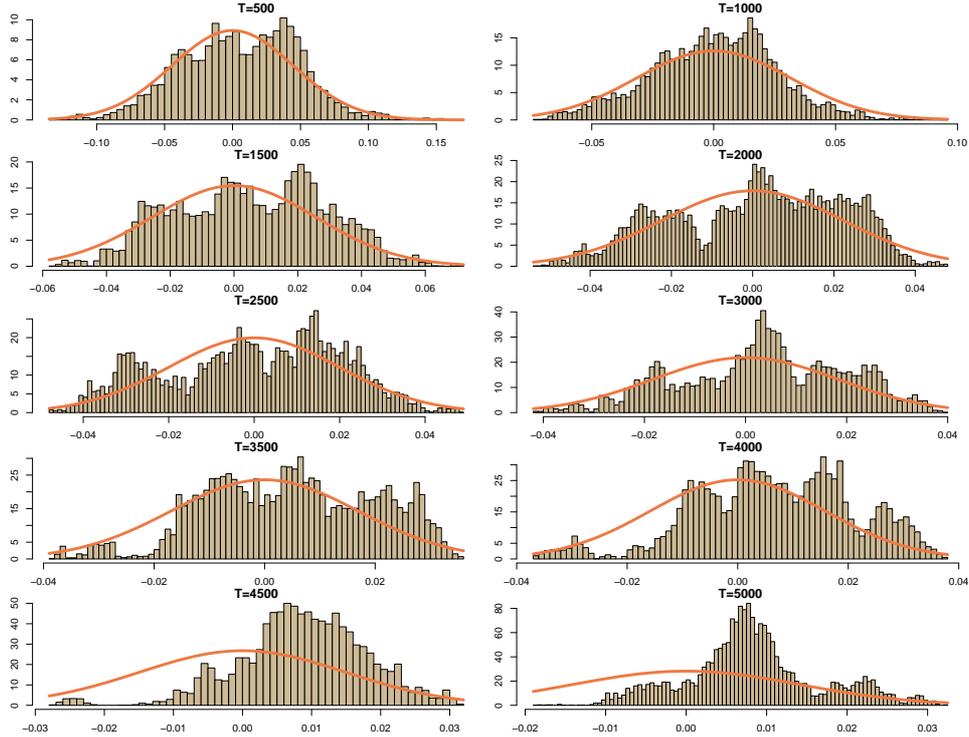}
\end{center}
\caption{\label{fig:compa}
Evolution of the particle learning sample against the target distribution
in terms of the number $T=50,\ldots,5000$ of iterations, for a particle population of fixed size $10^4$.}
\end{figure}

\begin{figure}
\begin{center}
 \includegraphics[height=0.5\textwidth,angle=270]{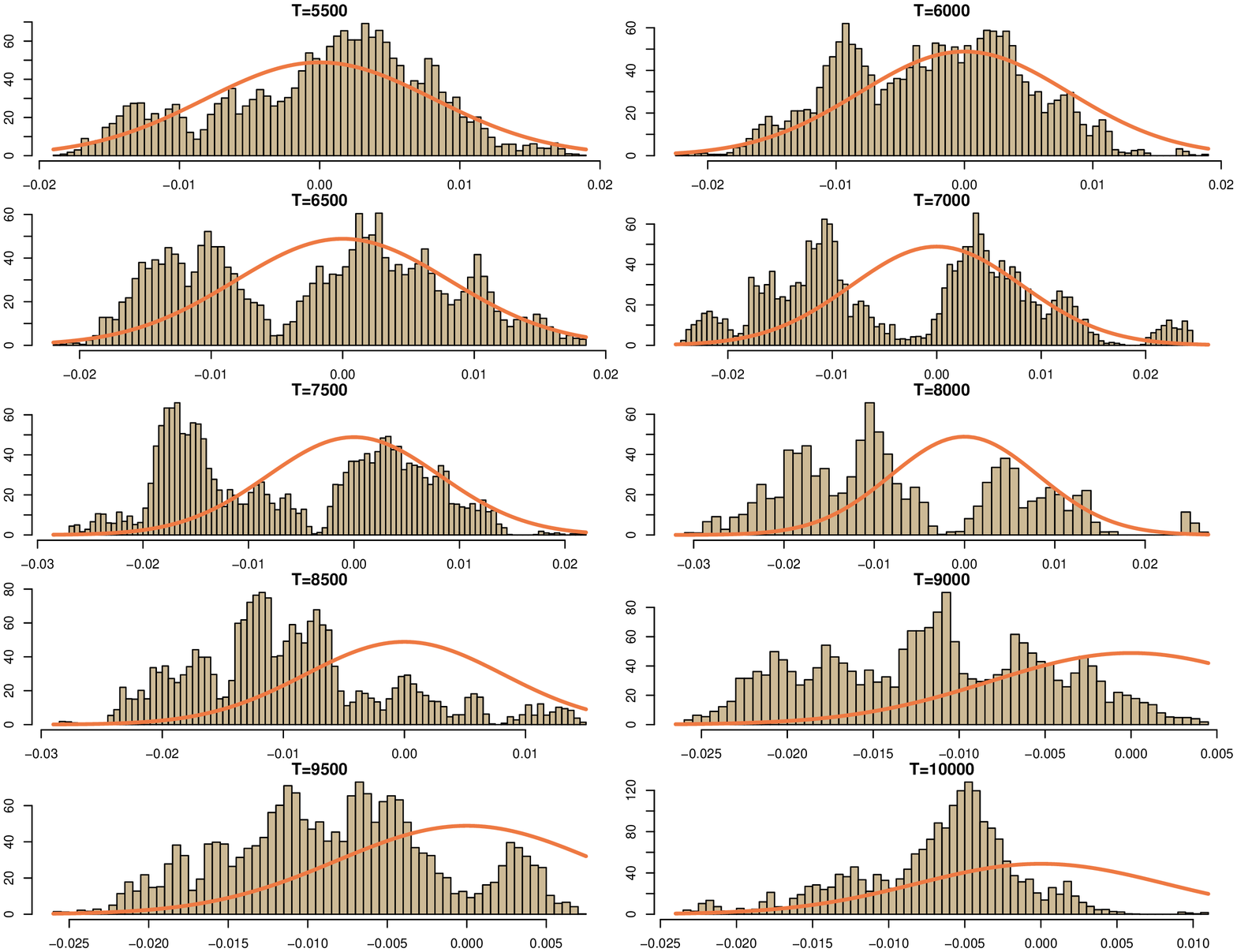}\includegraphics[height=0.5\textwidth,angle=270]{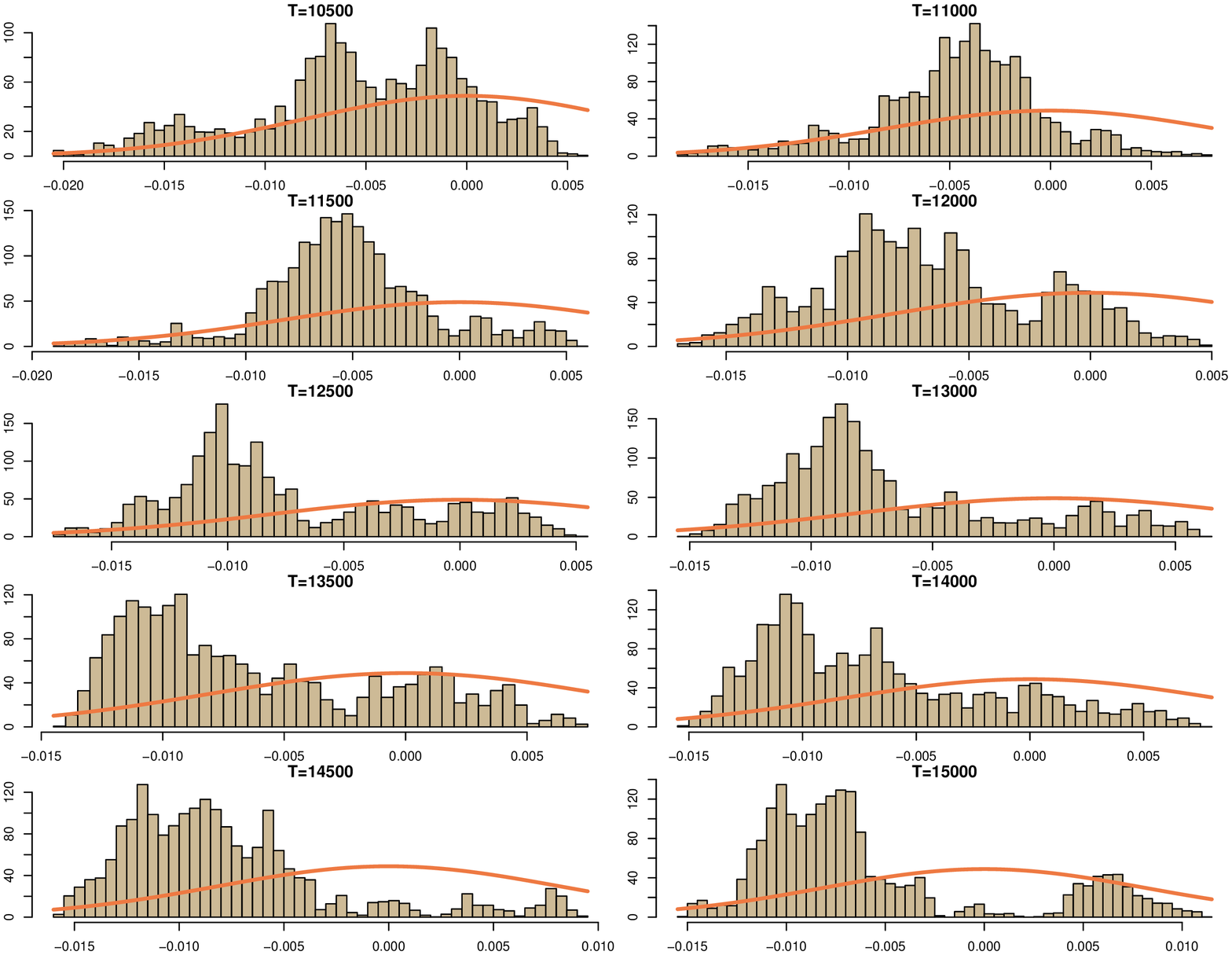}
\end{center}
\caption{\label{fig:compa3}
Figure \ref{fig:compa} continued for $T=5000,\ldots,15000$ iterations.}
\end{figure}

\locsection{A simple particle learning example}
Consider the ultimate case where the $z_t$'s are completely independent from the observations $y_t$, 
$z_t\sim\mathcal{N}(0,1)$, and where the empirical average of the $z_t$'s is the sufficient statistic. 
In this setting, the PL algorithm simplifies into the following iteration $t$:
\begin{enumerate}
\item Resample uniformly from $(Z_1^t,\ldots,Z_n^t)$ to produce $(\ev_1^t,\ldots,\ev_n^t)$;
\item Generate $z_{it}\sim\mathcal{N}(0,1)$;
\item Update $Z_i^{t+1} = (t\ev_i^t+z_{it})/(t+1)$
\end{enumerate}
The target distribution of the (sufficient) empirical average
$$
Z^t=(z_1+\ldots+z_t)/t
$$
is obviously the normal $\mathcal{N}(0,1/t)$ distribution. A straightforward simulation of the above particle system shows how
quickly the degeneracy occurs in the sample: Figures \ref{fig:compa}--\ref{fig:compa3} show a complete lack of fit to the 
target distribution as early as $t=500$ simulations when using $10,000$ particles.

\locsection{Conclusion}
The paper \cite{lopes:carvalho:johannes:polson:2010} fails to mention the well-documented issue of particle degeneracy 
\citep{cappe:moulines:ryden:2004,delmoral:doucet:jasra:2006}, thus giving the impression that PL escapes this problem. 
Our simple example shows that a particle system cannot be expected to withstand
an indeterminate increase in the number of observations without imposing a
corresponding exponential increase in the particle size.

\section{On the degeneracy of sufficient statistics (Robert, Ryder and Chopin)}
\bigskip
In connection with the discussion of Chopin and Robert, we detail in 
this discussion how the degeneracy dynamics of the particle learning technique 
of \cite{lopes:carvalho:johannes:polson:2010} impacts the distribution of 
the sufficient (or ``essential state vector") statistics.

\cite{lopes:carvalho:johannes:polson:2010} focus on the distribution of a sufficient statistic,
$p(Z_t|y^t)$, at time $t$. By insisting both on the low dimensionality of $Z_t$ and on the sufficiency,
they give the reader the impression that the poor approximation of the state vector $Z^t$ resulting from the 
resampling propagation scheme does not impact $p(Z_t|y^t)$, since their statement
{\em ``at time $T$, PL provides the filtered distribution of the last essential state vector $Z_T$, namely $p(Z_T|y^T)$"}
(Section 1.2) does not mention any deterioration in the approximation---this is how we understand {\em filtered}---provided 
by PL.  Because particle learning is inherently a particle filter \citep{pitt:shephard:1999},
this intuition is unfortunately wrong, as shown below in the case of an empirical 
average of the past auxiliary variables $Z_t$. Contrary to the belief that {\em ``resampling (...) is
fundamental in avoiding a decay"} (Section 1.2), resampling necessarily leads to degeneracy unless the size of the
particle population increases exponentially with $t$.

We thus consider again the case introduced by Chopin and Robert in their discussion, when the auxiliary variables
$z_t\sim\mathcal{N}(0,1)$ are independent from the observations $y_t$ 
and where the essential state vector statistic is the empirical average of the $z_t$'s. In this case, the
distribution of the empirical average
$$
Z^t=(z_1+\ldots+z_t)/t
$$
is the normal $\mathcal{N}(0,1/t)$ distribution, but the particle population degenerates into a single path
from the point of view of this sufficient statistic. In other words, degeneracy occurs much faster than the
root $T$ forgetting of the past of the particle path that is due to the averaging.
In order to support this perspective, we provide here a derivation of the variance of the particle population
after $t$ iterations. 

Using the same notations as in Chopin and Robert, since $\mathbb{E}[Z_i^t]=0$, $\text{var}(Z_i^t)=1/t$ and $Z_i^t = \frac{t-1}{t}\ev_i^t+\frac{z_{it}}{t}$, we consider
\begin{align*}
\BE[Z_i^t Z_j^t] &=\left(\frac{t-1}{t}\right)^2 \BE[\ev_i^t\ev_j^t] \\
&= \left(\frac{t-1}{t}\right)^2 \left(\BP[\ev_i^{t-1}=\ev_j^{t-1}]\BE[(\ev_i^{t-1})^2] + \BP[\ev_i^{t-1}\ne\ev_j^{t-1}]\BE[\ev_i^{t-1}\ev_j^{t-1}]\right)\\
&= \left(\frac{t-1}{t}\right)^2 \left(\frac{1}{n}\frac{1}{t-1} + \frac{n-1}{n}\BE[Z_i^{t-1}Z_j^{t-1}]\right).
\end{align*}

Now let $u_t = t^2 \BE[Z_i^t Z_j^t] -t+n$. The last line becomes $u_t=\frac{n-1}{n}u_{t-1}$. Since $u_1=n-1$, we have
\begin{align*}
\BE[Z_i^t Z_j^t] &= \frac{u_t+t-n}{t^2} =  \frac{\left(\frac{n-1}{n}\right)^{t-1}(n-1)+t-n}{t^2}\\
&=\frac{1}{t^2 n^{t-1}}\,\left\{(n-1)^t-n^t+tn^{t-1}\right\} = \frac{t-1}{2t}\,\frac{n^{t-2}}{n^{t-1}}+\cdots= \mathrm{O}_n(n^{-1})\,.
\end{align*}
In conclusion,
\begin{align*}
\text{var}(\overline Z^t_i) 
&= \frac{1}{nt}+\frac{n(n-1)}{n^2}\,\frac{1}{t^2 n^{t-1}}\,\left\{(n-1)^t-n^t+tn^{t-1}\right\}\\
&= \frac{1}{nt}\,\left[ 1+\frac{n(n-1)}{t}\,\left\{(1-1/n)^t-1+t/n\right\}\right].
\end{align*}

For $n$ fixed, and $t\rightarrow +\infty$, $t\text{var}(\overline
Z^t_i) \rightarrow 1$, a limit that does not depend on $n$, i.e. the
system eventually degenerates to a single path.  If we set $n=ct$,
then $nt\text{var}(\overline Z^t_i) \rightarrow C$, for some
$C>0$. Bearing in mind that the actual posterior variance should be
$O(t^{-1})$, this means that, to bound the {\em relative error}
uniformly over a given time interval, i.e. for $t=1,\ldots T$, one must take $n=O(T)$.

\section{On the degeneracy of path functionals in SMC (Chopin and Sch\"{a}fer)}
\bigskip
Much of the confusion around the degeneracy of particle learning and
similar algorithms \citep{fearnhead:2002,storvik:2002} seems related
to the lack of formal results regarding the degeneracy of path
functional in Sequential Monte Carlo. We'd like to report here some
preliminary investigation on this subject.

Consider a standard state-space model, with observed process $(y_t)$,
and hidden Markov process $(x_t)$, and a basic particle filter, which
would track the complete trajectory $x_{1:t}$, i.e. which would
produce, at each iteration $t$, $N$ simulated trajectories
$x_{1:t}^{(n)}$, with some weight $w_{t}^{(n)}$, so as to approximate
$p(x_{1:t}|y_{1:t})$. It is well known that the Monte Carlo error
regarding the expectation of $\varphi(x_{1:t})$ (a) remains bounded
over time if $\varphi(x_{1:t})=x_t$, (the filtering problem), and (b)
blows away, at an exponential rate, if $\varphi(x_{1:t})=x_1$ (the
smoothing problem). Chopin (2004) formalises these two statements by
studying the asymptotic variance that appears in the central limit
theorem for the corresponding particle estimates.

As mentioned above, and to the best of our knowledge, there is
currently no formal result on the divergence of the
asymptotic variance for test functions like $\varphi(x_{1:t})=t^{-1}\sum_{i=1}^t x_i$,
 i.e. some symmetric function
with respect to the complete trajectory. (The fact that this function
is a sufficient statistic should not play any role in this convergence
study.)  One difficulty is that the iterative definition of the
asymptotic variance given by \cite{chopin:2003} leads to cumbersome calculations. 

We managed however to compute this asymptotic variance exactly, for
the Gaussian local level model:
$$ x_{t+1}\mid x_t \sim N(x_t,1),\quad y_t\mid x_t \sim N(x_t,1)$$
and the functional $\varphi(x_{1:t})=t^{-1}\sum_{i=1}^t x_i$. In this case,
the asymptotic variance diverges at rate $O(e^{ct}/t^{2})$. Exact
calculations may be requested from the authors. We plan to extend
these results to a slightly more general model, e.g. with unknown
variances, and a function $\varphi$ which would be a sufficient
statistic for such parameters. We conjecture that this exponential divergence
occurs for many models: basically, in an average like
$\varphi(x_{1:t})=t^{-1}\sum_{i=1}^t x_i$, the Monte Carlo error attached
to $x_1/t$ should be $O(e^{ct}/t^2)$, and should dominate all the other
terms. This is at least what one observes in toy examples.
After, say, 100 iterations of a particle filter, the number of
distinct values within all the simulated trajectories (that have
survived so far)  for the component $x_1$ is typically very small, and
the degeneracy in the $x_1$ dimension seems sufficient to endanger the 
accuracy of any estimate based on the complete trajectory $x_{1:t}$.

\section{Remarks on the rejoinder (Robert)}
\newenvironment{kvothe}{\small\begin{quote}\begin{em}}{\end{em}\end{quote}\normalsize\vglue -0.2truecm}

Lopes et al. (2010) published a rejoinder on the discussions of their paper and the
following is a detailed examination of the arguments found in this rejoinder,
which requires a preliminary reading of the above papers as well as our
discussion. (All quotes are taken verbatim from the rejoinder.)

\begin{kvothe}``Particle learning based on the product estimate and MCMC
based on Chib's formula produce relatively similar results either for small or
large samples."\end{kvothe}

This statement about the estimation of the marginal likelihood (or the
evidence) and the example A that is associated with it in the rejoinder thus
comes to contradict our (rather intensive) simulation experiment which, as
reported in the discussion (Section \ref{sec:vid}), concludes to a strong bias
in evidence approximation induced by using particle learning, whether or not
the product estimator is used. We observed there that there were two levels of
degeneracy, one due to the product solution (errors in a product being more
prone to go and multiply) and one due to the particle nature of the sequential
method (which does not refresh particles from earlier periods). Figures
\ref{fig:kompa1}--\ref{fig:kompa4} are at odds with the one presented in the
rejoinder, maybe because we consider $10,000$ observations rather than $100$.
(I also fail to understand how the ``Log-predictive (TRUE)" quantity is
derived.)

\begin{kvothe}``Black-box sequential importance sampling algorithms and related
central limit theorems are of little use in practice."\end{kvothe}

This is a quote from the rejoinder that is rather puzzling. There is nothing
wrong with the central limit theorem which is the basis of error assessment in
Monte Carlo studies \citep{robert:casella:2009}. Indeed, one major consequence
of the central limit theorem is that it provides a precise scale for the speed
of convergence of Monte Carlo estimates and thus an indicator on the number of
particles needed for a given precision level. The authors of the rejoinder then
criticise our use of ``$1000$ particles in $5000$ dimensional problems" as we
``shouldn't be surprised at all with some of our findings". This is not
factually exact since I find no trace in the discussion of such a case: we use
$10,000$ particles in all examples and the target is either the distribution of
the 4 mixture parameters, the evidence  or the distribution of a
one-dimensional sufficient statistic.  Furthermore, these values of n and N are
those used in their example D. More importantly, nor the paper neither the
rejoinder map a practical strategy on how to increase the computational effort
along with the number of observations.

    \begin{kvothe}``This argument [that the Monte Carlo variance will `blow-up']
is incorrect and extremely misleading."\end{kvothe}

This point is central both to the discussions above and to the rejoinder, as
the authors maintain that the inevitable particle degeneracy does not impact
the distribution of the sufficient statistics. The argument about using time
averages over particle paths rather than sums appears reasonable at first.
Actually, taking an empirical average in almost stationary situations should
produce an approximately normal distribution. With an asymptotic variance
different from 0 (thanks to the central limit theorem) However, this is not the
main argument used in the discussions. Degeneracy in the particle path means
that the early terms in the average are less and less diverse in the sample
average. Therefore it is not that surprising that the variance is decreasing
down to too small a value! As shown in Figure \ref{fig:compa} above, degeneracy
due to resampling may induce severe biases in the distribution of empirical
averages while giving the impression of less variability (which is a recurrent
argument in the rejoinder). Furthermore, the fact that parameters are simulated
[rather than fixed] in the particle filter means that the process is not
geometrically ergodic, hence that Monte Carlo errors tend to accumulate along
iterations, rather than compensate. (This is why the comparison between PL and
sampling importance resampling is particularly relevant, because it does not
address this accumulation.) The rejoinder also quotes
\cite{olsson:cappe:douc:moulines:2008} for justifying the decrease in the Monte
Carlo variance. This is somehow surprising in that (a)
\cite{olsson:cappe:douc:moulines:2008} show that there is degeneracy without a
fixed-lag smoothing and (b) they require a geometric forgetting property on the
filtering dynamics. In addition, I think that Example E used to illustrate the
point about variance reduction is not very appropriate for this issue because
the hidden Markov chain is a Gaussian random walk, hence cannot be stationary
(a fact noted by the authors). And once again a decrease in the ``MC error"
does not mean a converging algorithm because degeneracy naturally induces
empirical variance decrease. (I also fail to see why the ``prior" on $(x_t)$ is
improper.) The final (if recurrent) argument that ``PL parameters do not
degenerate" is somehow puzzling: by nature, those parameters are simulated from
a distribution conditional on the sufficient parameters. So obviously the
simulated parameters all differ.  But this does not mean that they are
marginally distributed from the right distribution.

    \begin{kvothe}``MCMC schemes depend upon the not so trivial task of assessing convergence.
How long should the burn-in $G_0$ be?"\end{kvothe}

The rejoinder concludes with recommendations that sound more like a drafted
to-do note the authors forgot to remove than an accumulation of true
recommendations. (The above quote rather clearly supports our first point in
the discussion.) It seems to me that the comparison between MCMC and particle
filters is not particularly relevant, simply because particle filters apply in
[sequential] settings where MCMC cannot be implemented. To try to promote PL
over MCM by arguing that MCMC produces dependent draws while having convergence
troubles is not needed (besides, PL also produces [unconditional] dependent
draws). To advance that the Monte Carlo error for PL is of orser $C_T/\sqrt{N}$ is not
relevant either because $C_T$ is exponential in $T$ and because MCMC also has an
error in $\sqrt{N}$.

\bibliography{biblio}
\bibliographystyle{ba}

\end{document}